\documentclass[fleqn,usenatbib]{mnras}
\usepackage{graphicx}
\usepackage{newtxtext,newtxmath}
\usepackage[T1]{fontenc}
\usepackage{savesym}
\savesymbol{leftrotoavesymbol{uproot}}
\savesymbol{iint}
\savesymbol{iiint}
\savesymbol{iiiint}
\savesymbol{idotsint}
\savesymbol{Bbbk}
\usepackage{amsmath}	
\usepackage{array}
\usepackage{bm}
\usepackage[dvipsnames,svgnames]{xcolor}
\usepackage{soul}
\usepackage[ruled,lined,linesnumbered]{algorithm2e}
\usepackage{amssymb,physics,esint}
\usepackage{multirow}

\DeclareRobustCommand{\VAN}[3]{#2}
\let\VANthebibliography\thebibliography
\def\thebibliography{\DeclareRobustCommand{\VAN}[3]{##3}\VANthebibliography}

\title[A High Order FPFS Shear Estimator]{Accurate Shear Estimation with Fourth-Order Moments}

\author[Park et al.]{
Andy Park,$^{1}$\thanks{E-mail: chanhyup@andrew.cmu.edu}
Xiangchong Li,$^{1, 2}$
Rachel Mandelbaum$^{1}$
\\
$^{1}$McWilliams Center for Cosmology and Astrophysics, Department of Physics, Carnegie Mellon University, 5000 Forbes Ave, Pittsburgh, PA 15213\\
$^{2}$Brookhaven National Laboratory, Bldg 510, Upton, New York 11973, USA
}

\date{Accepted XXX. Received YYY; in original form ZZZ}

\pubyear{2023}

\begin{document}
\label{firstpage}
\pagerange{\pageref{firstpage}--\pageref{lastpage}}
\maketitle

\begin{abstract}
As imaging surveys progress in exploring the large-scale structure of the Universe through the use of weak gravitational lensing, achieving subpercent accuracy in estimating shape distortions caused by lensing, or shear, is imperative for precision cosmology. 
In this paper, we extend the \texttt{FPFS} shear estimator using fourth-order shapelet moments and combine it with the original second-order shear estimator to reduce galaxy shape noise. We calibrate this novel shear estimator analytically to a subpercent level accuracy using the \texttt{AnaCal} framework. This higher-order shear estimator is tested with realistic image simulations, and after analytical correction for the detection/selection bias and noise bias,  
the multiplicative shear bias $|m|$ is below $3\times10^{-3}$ ($99.7\%$ confidence interval) for both isolated and blended galaxies. Once combined with the second-order \texttt{FPFS} shear estimator, the shape noise is reduced by $\sim35\%$ for isolated galaxies in simulations with HSC and LSST observational conditions. However, for blended galaxies, the effective number density does not significantly improve with the combination of the two estimators. Based on these results, we recommend exploration of how this framework can further reduce the systematic uncertainties in shear due to PSF leakage and modelling error, and potentially provide improved precision in shear inference in high-resolution space-based images. 
\end{abstract}

\begin{keywords}
gravitational lensing: weak; cosmology: observations; techniques: image processing.
\end{keywords}



\section{Introduction}



Weak gravitational lensing refers to the small but coherent distortions of distant galaxies' light profiles due to intervening massive foreground matter between the source galaxies and observers. Analyzing the statistics of this distortion can reveal the distribution of matter and the large-scale structure, yielding deep insights into the fundamental physics of the Universe and its evolution (see \citealt{kilbinger} for a review of weak lensing). Measuring the coherent pattern of distortions, or shear, in the observed image is one of the most effective ways to constrain the fundamental physics of the Universe. Upcoming Stage IV imaging surveys like the Vera C.\ Rubin Observatory Legacy Survey of Space and Time \citep[LSST;][]{2009arXiv0912.0201L, 2019ApJ...873..111I}, \textit{Euclid} \citep{2011arXiv1110.3193L}, and \textit{Nancy Grace Roman} Space Telescope High Latitude Imaging Survey \citep{2019arXiv190205569A} will cover a large sky area and observe more than billion source galaxies, allowing us to make unprecedented percent-level cosmic shear measurements.

However, the magnitude of this tiny distortion, which causes percent-level changes in the ellipticities of the observed galaxy images, is on average $\sim$$10\%$ of the root mean square (RMS) of intrinsic galaxy shapes. Hence an accurate measurement of weak lensing shear is complex and needs to calibrate several sources of systematic effects that complicate the process of characterizing galaxy shapes, including the point-spread function (PSF) from atmospheric effects and telescope optics \citep[e.g.,][]{Liaudat2023}; noise bias from image noise due to the non-linearity in shear estimator \citep[e.g.,][]{image_noise_bias}; model bias due to unrealistic assumptions about galaxy morphology \citep[e.g.,][]{Gary2010}; selection bias due to sample selection \citep{selection_bias} and detection \citep{metaDet_Sheldon2020}; and biases from blending and deblending of galaxy light profiles \citep{fpfs1}. To achieve percent-level bias in cosmological parameters despite systematics 
(see \citealt{2018ARA&A..56..393M} for a review of systematics in weak lensing), the upcoming surveys require that the residual systematics after all corrections should be well below the statistical uncertainty.  This requires that  residual systematic uncertainties in shear measurements be below one part per thousand \citep{2018arXiv180901669T}.

Several recent shear measurement techniques have been developed that aim to reduce the reliance on calibration using external simulations.  This is desirable because the simulations may not be sufficiently realistic to achieve the level of precision needed for cosmological inference. These methods that aim for unbiased shear inference include \textsc{Metadetection} \citep{metaDet_LSST2023}, a numerical self-calibration method; and \texttt{BFD} \citep{BFD_Berinstein2016}, a Bayesian approach to shear estimation. There has been significant efforts to develop purely analytical shear estimators \citep{fpfs1, fpfs2, fpfs, fpfs3, fpfs4} in order to achieve subpercent accuracy without relying on any calibration from external image simulations. The Fourier power function shapelets method (\texttt{FPFS} hereinafter) can achieve a sub-percent accuracy by correcting for noise bias, selection and detection bias. 
In short, \texttt{FPFS} uses set of shapelet modes \citep{shapeletsI_Refregier2003, polar_shapelets_Massey2005, BernsteinSSSS}, detection modes and other linear observables after PSF deconvolution to carry out the detection, selection, and measurement of the galaxy ellipticity and response to shear. It uses the first-order derivatives (Jacobian) of the ellipticity with respect to these linear observables and the shear responses of shapelet to obtain the shear response of the ellipticity. To correct for the noise bias, one can use the second-order derivative (Hessian matrix) of the ellipticity and the covariance matrix of the measurement error on the observables \citep{fpfs3}, or add pure image noise to the galaxy image with a carefully-chosen noise correlation function to derive an analytical noise bias correction \citep{fpfs4}. This innovative analytical calibration technique for shear estimation, designated \texttt{AnaCal} \citep{fpfs, fpfs3} is more than a hundred times faster than the current benchmark \textsc{Metadetection} algorithm, documented in \citet{metaDet_LSST2023}. 

Lensing shear causes the estimates of the ellipticities of distant galaxies to change. This effect is quantified statistically by observing the characteristics of a galaxy that undergoes simple transformations under shear. One way to parameterize the shape is based on second-order moments of the galaxy image, which captures the orientation and elongation of the object. These moments have been widely adopted in weak lensing studies due to their robustness and relative insensitivity to image noise. However, higher-order moments, while more sensitive to image noise and harder to model, offer the potential to capture additional information about the galaxy's response to shear. They probe finer structural details that are not accessible through second moments alone, suggesting they may contain complementary information about the shear. 
The original \texttt{FPFS} implementation primarily focused on second-order shapelet moments to construct the ellipticity of galaxies, and achieves a shear estimation bias below $0.3\%$ in the presence of blending. In this work, we extend the \texttt{FPFS} framework by incorporating fourth-order shapelet moments to define the ellipticity.  We use image simulations to test the accuracy of our new shear estimator after analytically correcting for detection and selection bias. We then take two \texttt{FPFS} shear estimators, second- and fourth-order, and combine them to maximize the shear signal and reduce the overall shape noise. Our goal is to quantify how much additional information is provided by the fourth-order moments; the fourth-order shear estimator, which is independent of the second-order, can also be used to cross-comparison and improve the systematic control.

This paper is organized as follows. In Section~\ref{sec:method}, we give a brief overview of the \texttt{FPFS} shear estimator within the \texttt{AnaCal} framework, the fourth-order shear estimator, and introduce a method to combine two different shear estimators. In Section~\ref{sec:simulation}, we present galaxy image simulations that we used to test the accuracy of our new fourth-order shear estimator. In Section~\ref{sec:result}, we show the result of our analysis and quantify the reduction of shape noise by combining two shear estimators. Finally, in Section~\ref{sec:conclusion}, we summarize our results and future outlook.

\section{Method}
\label{sec:method}



In this section, we briefly review the \texttt{FPFS} shear estimator developed in \citet{fpfs1, fpfs2} 
and calibrated with the \texttt{AnaCal} framework implemented in \citet{fpfs, fpfs3}. We then extend \texttt{FPFS} to a higher-order shear estimator.

The distortion of galaxy shapes, or shear, caused by foreground inhomogeneous mass distribution can be described by a locally linear transformation (or the Jacobian matrix) as
\begin{equation}
    A = \begin{pmatrix}
        1 - \gamma_1 & -\gamma_2 \\
        -\gamma_2 & 1 + \gamma_1
    \end{pmatrix},
\end{equation}
where the component $\gamma_1$ quantifies the amount of stretching of the image along the horizontal direction and $\gamma_2$ quantifies the stretching of the image in the direction at an angle of 45 deg with the horizontal direction. We use a complex spinor to represent shear as $\gamma = \gamma_1 + \mathrm{i} \gamma_2$, where $\mathrm{i}$ is the imaginary number unit. In this paper, we set the lensing convergence to zero to simplify the notation. The shear is typically estimated by measuring the galaxy's ellipticity, a spin-2 observable, which negates under a 90-degree rotation (see Appendix B of \citealt{fpfs} for more details). In the weak lensing limit, shear is on the order of a few percent or less ($\abs{\gamma} \lesssim 0.02$),
making the shear signal much smaller than the shape noise due to galaxy intrinsic shapes, so that a large ensemble of galaxies is needed to infer shear.

\subsection{Galaxy Detection}
\label{sec:galaxy_detection}

Before measuring shear, detection and selection of galaxies can introduce biases that affect the shear estimation, hence it is essential to derive their shear response for an accurate shear estimation. Galaxy detection from images in \texttt{FPFS} uses four detection modes ($\nu_i$, where $i=0\dots 3$) for each pixel in the image to characterize the difference in the value of the pixels with respect to the nearby pixels in four directions. These nearby pixel detection modes are then used to identify peaks that are served as ``peak candidates''.  
The corresponding selection bias from carrying out the detection process using these peak modes is analytically corrected using the shear responses of the pixel values \citep{fpfs2, fpfs4}. For a galaxy profile $f(\bm{x})$, with $\bm{x}$ denoting image in real space, we define the detection modes for every pixel $i$ as 
\begin{equation}
    \label{eq:detection_modes}
    \nu_i = \iint \mathrm{d}^2k \: \psi^*_i(\bm{k}) \frac{f^p(\bm{k})}{p(\bm{k})},
\end{equation}
where $f^p(\bm{k})$ is the observed (PSF-convolved, noisy) image in Fourier space and $p(\bm{k})$ is the PSF image in Fourier space, and the coordinate center is set to the center of this pixel. The detection kernels $\psi^*_i(\bm{k})$ for wave number vector $\bm{k} = (k_1, k_2)$ are defined in Fourier space as
\begin{equation}
    \psi^*_i(\bm{k}) = \frac{1}{(2\pi)^2}e^{-|\bm{k}|^2\sigma_h^2/2}(1 - e^{\mathrm{i}\bm{k}\cdot\bm{x}_i}),
\end{equation}
where $\bm{x}_i = (x_i, y_i) = (\cos(i\pi/2), \sin(i\pi/2)$ are position vectors to nearby pixels with lengths equal to the image pixel side length and orientations pointing towards the four directions separated by $\pi/2$. 
The shear response of these detection modes is given by
\begin{equation}
    \nu_{i;\alpha} = \iint \mathrm{d}^2 k \: \psi^*_{i;\alpha}\frac{f^p(\bm{k})}{p(\bm{k})},
\end{equation}
where the subscript ``$;\alpha$'' denotes the partial derivative with respect to one component of the shear, $\gamma_\alpha$. The shear response of each detection kernel can be written as a combination of shapelet basis and is given by
\begin{align}
    \begin{split}
        \psi_{i;1}^* \equiv \frac{\partial \psi_i^*}{\partial\gamma_1} &= \frac{1}{(2\pi)^2}e^{|\bm{k}|^2\sigma_h^2/2}(k_1^2-k_2^2)\sigma_h^2\left(1 - e^{\mathrm{i}(k_1x_i + k_2y_i)}\right) \\
        &-\frac{1}{(2\pi)^2}e^{|\bm{k}|^2\sigma_h^2/2}(\mathrm{i}x_ik_1 - \mathrm{i}y_ik_2)e^{\mathrm{i}(k_1x_i + k_2y_i)},
    \end{split}\\
    \begin{split}
        \psi_{i;2}^* \equiv \frac{\partial \psi_i^*}{\partial\gamma_2} &= \frac{1}{(2\pi)^2}e^{|\bm{k}|^2\sigma_h^2/2}(2k_1k_2)\sigma_h^2\left(1 - e^{\mathrm{i}(k_1x_i + k_2y_i)}\right) \\
        &-\frac{1}{(2\pi)^2}e^{|\bm{k}|^2\sigma_h^2/2}(\mathrm{i}y_ik_1 + \mathrm{i}x_ik_2)e^{\mathrm{i}(k_1x_i + k_2y_i)}.
    \end{split}
\end{align}

\subsection{\texttt{FPFS} Shapelet Modes}
\label{sec:FPFS_shapelet_modes}
\texttt{FPFS} uses polar shapelet modes \citep{polar_shapelets_Massey2005} to construct various galaxy properties, including flux, size, and shape. These polar shapelet modes are constructed by projecting the observed noisy galaxy image after PSF deconvolution onto a set of Gaussian-weighted orthogonal functions \citep{fpfs1}. The \texttt{FPFS} complex polar shapelet modes are defined as:
\begin{equation}
\label{eq:shapelets_modes}
    M_{nm} \equiv \iint \mathrm{d}^2k \, \tilde{\chi}^{*}_{nm}(\bm{k}) \frac{f^p(\bm{k})}{p(\bm{k})}\, ,
\end{equation}
where $\tilde{\chi}_{nm}$ is the Fourier transform of the polar shapelet basis function characterized by a radial quantum number ``$n$" and an angular quantum number, or spin number, ``$m$". Polar shapelet basis functions $\chi_{nm}$ are defined as
\begin{equation}
\label{eq:shapeletmodes}
\begin{split}
\chi_{nm}(\bm{x} \,|\, \sigma_h)&=(-1)^{(n-|m|)/2}\left\lbrace
    \frac{[(n-|m|)/2]!}{[(n+|m|)/2]!}\right\rbrace^\frac{1}{2}\\
    &\times
    \left(\frac{\absolutevalue{\bm{x}}}{\sigma_h}\right)^{|m|}
    L^{|m|}_{\frac{n-|m|}{2}}\left(\frac{\absolutevalue{\bm{x}}^2}{\sigma_h^2}\right)
    e^{-\absolutevalue{\bm{x}}^2/2\sigma_h^2}e^{-im\theta},
\end{split}
\end{equation}
where $L^{|m|}_{\frac{n-|m|}{2}}$ are the Laguerre polynomials and $\sigma_h$ is the smoothing scale of shapelets and the detection kernel. $n$ can be any nonnegative integer and $m$ is an integer between $-n$ and $n$ in steps of two.

Under a Fourier transform, the shapelet basis changes as
\begin{equation}
    \chi_{nm}(\bm{x} | \sigma_h) \to \tilde{\chi}_{nm}(\bm{k}) = i^n \chi_{nm}(\bm{k} | 1/\sigma_h),
\end{equation}
so the amplitude of the shapelet basis in Fourier space is the same as in real space but the scale is the inverse of that in real space \citep{shapeletsI_Refregier2003}. Typically, $\sigma_h$ is set so that it is greater than the scale radius of the PSF in configuration space so that the deconvolution does not amplify the noise on small scales, or large $|k|$. In this work, we do not adapt $\sigma_h$ to the size of the galaxy light profile. However, in real observations, we can split the survey into smaller patches and have a different smoothing scale $\sigma_h$ for each patch.

When a galaxy image is distorted under the influence of shear, a finite number of independent shapelet modes are coupled (separated by $|\Delta n| = 2$  and $|\Delta m| = 2$). This causes the sheared shapelet modes to be a linear combination of a finite number of other shapelet modes. Since the \texttt{FPFS} shapelet modes are computed after deconvolving the galaxy with the PSF model, the PSFs do not bias shear estimation as long as we have a good PSF model at the positions of the galaxies.

\begin{figure}
    \centering
    \includegraphics[width=0.5\textwidth]{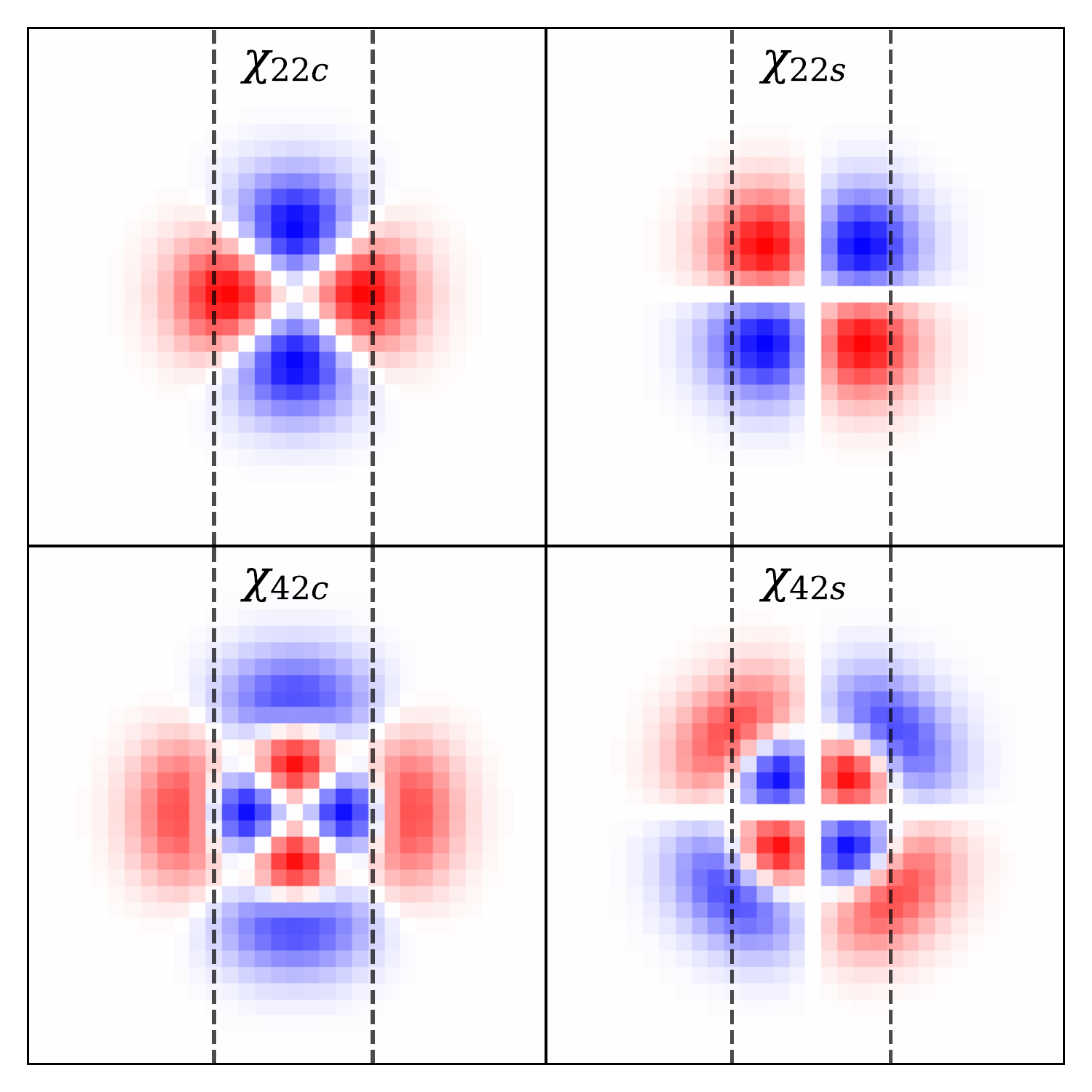}
    \caption{The real and imaginary components of the spin-2 second order and fourth-order shapelet basis. The fourth-order shapelets are sensitive to scales larger and smaller than that of the second moments, as referenced by the dashed lines. The color scale assigned to each basis function spans the interval $[-A, A]$, with $A$ representing the maximum absolute value of the corresponding basis function.
    }
    \label{fig:shapelet}
\end{figure}

\subsubsection{Galaxy Flux and Galaxy Size}
The zeroth order shapelet mode, $M_{00}$, is the value of the central peak of the smoothed image with the smoothing scale of $\sigma_h$. We follow \citet{fpfs} to use $M_{00}$ to quantify the brightness of galaxies, and \texttt{FPFS} flux is defined as
\begin{equation}
    F = \frac{M_{00}}{\iint \mathrm{d}^2k |\tilde{\chi}_{00}(\bm{k})|^2/|p(\bm{k})|^2},
\end{equation}
where the denominator is the square of the $L^2$ norm of the resmoothing kernel. We can use the flux and convert it to \texttt{FPFS} magnitude defined as
\begin{equation}
    m_\textrm{F} = m_\textrm{zero} - 2.5 \log(F),
\end{equation}
where $m_\textrm{zero}$ is the zero point of the survey. The value of $m_\textrm{zero}$ is 27 for HSC coadded images \citep{HSCBosch2018} and 30 for LSST coadded images \citep{2019ApJ...873..111I}. 

The galaxy size is conventionally measured using second-order Gaussian weighted moments (see, e.g., \citealt{regauss}). Following \citet{fpfs}, we use the combination of spin-0 shapelet moments to define the size as
\begin{equation}
    \iint \mathrm{d}^2x f\left(\frac{|\bm{x}|}{\sigma_h}\right)^2 e^{-|\bm{x}|^2/2\sigma_h^2} = M_{00} + M_{20},
\end{equation}
and define the \texttt{FPFS} resolution as
\begin{equation}
    R_2 = \frac{M_{00} + M_{20}}{M_{00}}.
\end{equation}
Note that this definition of resolution given by \citet{fpfs} is conceptually similar to the resolution defined in \cite{regauss}. Under a shear distortion, the shapelet moments $M_{00}$ and $M_{20}$ change from their intrinsic values $\bar{M}_{00}$ and $\bar{M}_{20}$ as
\begin{align}
\nonumber
    M_{00} &= \bar{M}_{00} + \sqrt{2}(\gamma_1 \bar{M}_{22c} + \gamma_2 \bar{M}_{22s}),\\
    M_{20} &= \bar{M}_{20} + \sqrt{6}(\gamma_1 \bar{M}_{42c} + \gamma_2 \bar{M}_{42s}),
\end{align}
where we use $M_{nmc}$ and $M_{nms}$ to denote the real and imaginary part of the complex shapelet mode $M_{nm}$. We use their linear shear response to derive the shear response of the galaxy detection/selection and to correct for the detection/selection bias in Section~\ref{sec:weights}.

\subsection{Detection and Selection Weights}
\label{sec:weights}

We use the weight functions introduced in \citet{fpfs} for galaxy detection and selection. For the galaxy detection process, we apply cuts on peak modes, and for the galaxy selection process, we apply cuts on magnitude and resolution. The analytical shear responses of the hard cuts are noisy and unstable, especially when applying multiple cuts to the galaxy properties, as a hard selection weight is discontinuous and not differentiable at the selection boundary. 
For this work, instead of applying hard cuts on observables, we use truncated sine functions (see~equation 45 of \citealt{fpfs}) since these are differentiable and are more stable. The selection weight used to select the galaxy sample is given by
\begin{equation}
    \label{eq:selection_weight}
    w(\bm{\nu}) = T^{\text{sel}}_0(M_{00}) T^{\text{sel}}_2(M_{00} + M_{20}) \prod_{i=0}^3 T^{\text{det}}(\nu_i),
\end{equation}
where $T^{\text{sel}}_0$ is used to select bright galaxies with high signal-to-noise ratio, $T^{\text{sel}}_2$ is used to select well-resolved large galaxies, and $T^{\text{det}}$ is used to define the cut on peak modes. 
Since we have calculated the shear response of the detection modes (Section~\ref{sec:galaxy_detection}) and the shapelet modes used to quantify the flux and size (Section~\ref{sec:FPFS_shapelet_modes}), we can calculate the derivative of the detection and selection weight function to shear.

\begin{figure*}
    \centering
    \includegraphics[width=0.8\textwidth]{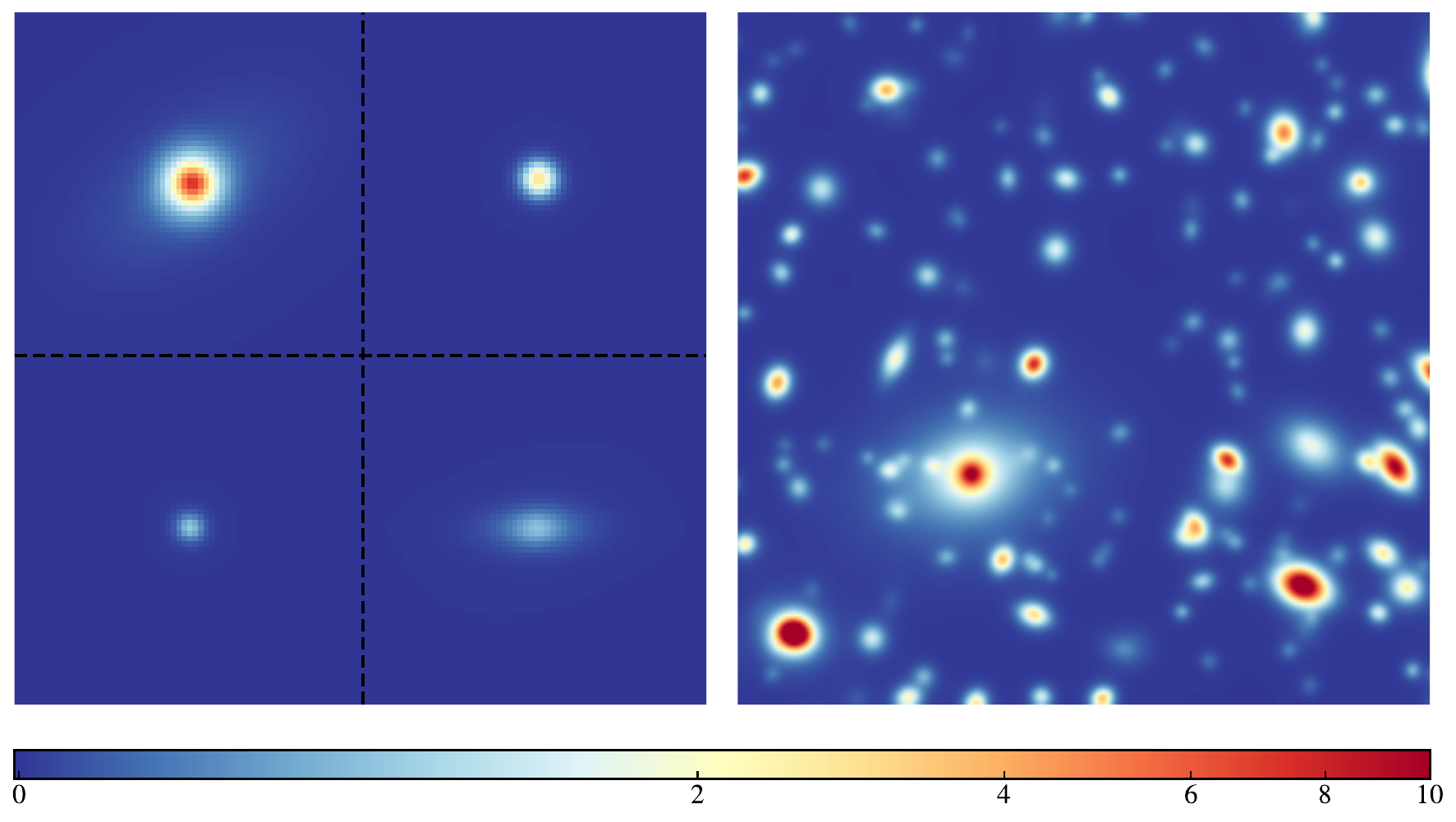}
    \caption{The left panel shows a 128 pix $\times$ 128 pix (equivalent to $0.36 \times 0.36$ arcmin$^2$) stamp image of the isolated galaxy image simulation with HSC seeing,  
    where the dotted black lines show the boundaries of the 64 pix $\times$ 64 pix stamps. The right panel shows a random cut-out coadded image of 240 pixels $\times$ 240 pixels (equivalent to $0.8 \times 0.8$ arcmin$^2$) of $griz-$bands of the LSST-like blended galaxy image simulation \citep{metaDet_LSST2023}. Both images were produced before adding noise. 
    }
    \label{fig:sim_img}
\end{figure*}

\subsection{Fourth-Order Shear Estimator}

\cite{fpfs} uses shapelet modes $M_{00}, M_{20}$, and $M_{22}$ (hereafter second order moments) and their shear responses to compute the galaxy flux, size, and shape (ellipticity). In this work, use the same detection and selection defined in \citet{fpfs, fpfs4} and extend the second-order shapes adopted by \citet{fpfs} to fourth-order shapes.


In Fig.~\ref{fig:shapelet}, we show the spin-2 second-moment and fourth-moment  basis functions. The fourth-moment basis is more sensitive to pixels with radius both larger and smaller than that of the second moment. Within the same polar angle, the sensitivities of fourth moments to smaller and larger radii have opposite signs, making $M_{42}$ sensitive to the difference in spin-2 behavior between pixels with small and large radii. Quantitatively, a shapelet model contains information only between the minimum and maximum scales \citep{polar_shapelets_Massey2005} defined as
\begin{equation}
    \label{eq:scale}
    \theta_\mathrm{min} = \frac{\sigma_h}{\sqrt{n+1}} \quad \mathrm{and} \quad \theta_\mathrm{max} = \sigma_h \sqrt{n+1},
\end{equation}
where $\sigma_h$ is the shapelets smoothing scale and $n$ is the shapelets order. We refer readers to Appendix~\ref{sec:apA} for a detailed analysis of the shapelet moments of a Gaussian profile and their sensitivity to shear.

Using the shear responses of shapelet modes defined in \cite{polar_shapelets_Massey2005}, the shapelet modes are given by
\begin{align}
    \begin{split}
    M_{42c} &= \bar{M}_{42c} + \frac{\sqrt{6}}{2} \gamma_1 \left(M_{20}
    - M_{60}\right)\label{eq:m42c} \\
    &- \sqrt{5} \gamma_1 M_{64c} - \gamma_2 \sqrt{5} M_{64s}
    \end{split}
     \\
     \begin{split}
     M_{42s} &= \bar{M}_{42s} + \frac{\sqrt{6}}{2} \gamma_2 \left(M_{20}
    - M_{60}\right)\label{eq:m42s}\\
    &+ \sqrt{5} \gamma_2 M_{64c} - \gamma_1 \sqrt{5} M_{64s},
     \end{split}
\end{align}
where $M_{nm}$ represents the sheared shapelet modes and $\bar{M}_{nm}$ represents the intrinsic shapelet modes in the absence of shear. As described in \cite{fpfs1}, shear can be inferred from equations~\eqref{eq:m42c} and \eqref{eq:m42s} by taking the expectation values on both sides. Assuming that there is no preferential direction of the randomly selected galaxy ensemble, the expectation values of the intrinsic spin-2 and spin-4 shapelet modes on the right-hand side of the equations reduce to zero, i.e. $\langle \bar{M}_{42}\rangle = \langle M_{64}\rangle = 0$. It is the population variance of these spin-2 and spin-4 shapelet modes that causes the shape noise of the shear estimator.

\cite{fpfs1} introduced a normalizing scheme to re-weight the shapelet modes dominated by bright galaxies and reduce the shape noise of the shear estimation. The dimensionless \texttt{FPFS} fourth-order ellipticity is defined as
\begin{equation}
    \label{eq:ellipticity_fourth}
    e_1 + \mathrm{i} e_2 = \frac{M_{42}}{M_{00} + C^{(4)}},
\end{equation}
In \cite{fpfs}, they use $M_{22}$ in the numerator of equation~\eqref{eq:ellipticity_fourth} to define the spin-2 ellipticity. In this work, we use $M_{42}$ to define the fourth-order spin-2 ellipticity. The \texttt{FPFS} weighting parameter $C^{(4)}$ adjusts the relative weight for galaxies with different brightness. Note that the optimal value of the weighting parameter $C^{(n)}$ is different for each order of the estimator, as different shapelet moments are sensitive to different scales. 

\subsection{Analytical Shear Calibration}
In this subsection, we outline how to correct for the noise bias in shear due to the pixel noise in the images. The previous generation of \texttt{AnaCal} corrected the noise bias by computing the second- and higher-order derivatives of the nonlinear observables \citep{fpfs3}. This required taking derivatives of smoothstep functions with smoothness parameters, 
which led to significant fluctuations in higher-order derivatives \citet{fpfs4}. \citet{Metacal2017} proposed a numerical recipe to correct for noise bias by adding additional noise to the already-noisy image. \citet{fpfs4} analytically prove the method is free of noise bias and adopts the analytical version within the \texttt{AnaCal} framework. This paper follows \citet{fpfs4} to analytically correct for noise bias in shear estimation. Following \citet{Metacal2017}, we introduce an additional layer of noise to the image that shares the same statistical properties after being rotated counterclockwise by $90^\circ$, with the rotation defined in the space prior to PSF convolution. The addition of this noise layer effectively cancels out the spin-2 anisotropies present in the original noise image after deconvolution. Note that \citet{Metacal2017} is a ``finite-difference'' version of the noise bias correction which is not the same as the ``analytical'' version in \citet{fpfs4}. This paper uses the ``analytical'' version, not the ``finite difference'' version. The shear response of the renoised ellipticity can be measured after adding the simulated noise as (adopting Einstein notation)
\begin{equation}
    \left<\tilde{\tilde{R}}_\alpha\right> =
    \left<\frac{\partial \left(\tilde{\tilde{w}} \tilde{\tilde{e}}_\alpha\right)}{\partial \tilde{\tilde{\nu}}_i}
    \left(\tilde{\tilde{\nu}}_{i;\alpha} - 2\delta\nu'_{i;\alpha}\right)\right>,
\end{equation}
where the subscript $\alpha$ denotes each component of the shear and $\delta\nu'_i$ is the measurement error of the $i$th linear observable from the additional noise. We use a double tilde $(\bm{\tilde{\tilde{\nu}}})$ to denote linear observables with doubled image noise. The shear estimator is then
\begin{equation}
    \hat{\gamma}_\alpha = \frac{\left<\tilde{\tilde{w}}\tilde{\tilde{e}}_\alpha\right>}{\left<\tilde{\tilde{R}}_\alpha\right>} + \mathcal{O}(\gamma^3),
\end{equation}
where $\tilde{\tilde{e}}$ is the ellipticity observed after adding the additional image noise. This shear estimator is free from noise bias and is accurate to second order of shear. It is worth noting that the \textit{renoising} approach does not require any computation of noisy second- and high-order derivatives, and does not include higher-order terms from image noise which are present in the original version of \texttt{FPFS}. However, a limitation of this method is that we need to double the image noise before running detection and source measurement. 
In this work, we used this \textit{renoising} approach to analytically correct for the noise bias.

\subsection{Combining Shear Estimators}
\label{sec:combine_estimators}
Given the second- and fourth-order shear estimators, we combine these estimators to maximize the SNR of the shear estimator and minimize the noise in the shear estimation, including intrinsic shape noise and measurement error from image noise. The combined shear estimate is defined as:
\begin{equation}
    \label{eq:shear_combined}
    \widehat{\gamma}_{\alpha,t} = \mu \widehat{\gamma}_{\alpha,2} + (1 - \mu)\widehat{\gamma}_{\alpha,4},
\end{equation}
where $\alpha$ is the component of the shear and the subscripts 2 and 4 are the order of the shear estimator. The statistical uncertainty in the combined shear estimation is then
\begin{align}
    \label{eq:shear_combined_error}
    \nonumber\operatorname{Var}(\widehat{\gamma}_{\alpha,t}) &= \mu^2 \sigma_{\gamma_{\alpha,2}}^2 + (1 - \mu)^2 \sigma^2_{\gamma_{\alpha,4}} \\
    &+2 \mu(1 - \mu) \operatorname{Cov}(\widehat{\gamma}_{\alpha,2}, \widehat{\gamma}_{\alpha,4}),
\end{align}
where $\sigma_{\gamma_\alpha}$ are the standard deviation of each estimator, and $\rho_\alpha$ is the correlation between the two estimators. The shear estimator weight $\mu$ is defined so that it minimizes the variance of the combined estimate as
\begin{align}
    \nonumber \mu &= \underset{\mu}{\arg\min} \operatorname{Var}(\widehat{\gamma}_{\alpha,t})\\
    \nonumber &= \frac{\sigma_{\gamma_{\alpha,4}}^2 - \operatorname{Cov}(\widehat{\gamma}_{\alpha,2}, \widehat{\gamma}_{\alpha,4})}{\sigma_{\gamma_{\alpha,2}}^2 + \sigma_{\gamma_{\alpha,4}}^2 - 2 \operatorname{Cov}(\widehat{\gamma}_{\alpha,2}, \widehat{\gamma}_{\alpha,4})} \\
    &= \frac{\sigma_{\gamma_{\alpha,4}}^2 - \rho_\alpha \sigma_{\gamma_{\alpha,2}} \sigma_{\gamma_{\alpha,4}}}{\sigma_{\gamma_{\alpha,2}}^2 + \sigma_{\gamma_{\alpha,4}}^2 - 2 \rho_\alpha \sigma_{\gamma_{\alpha,2}} \sigma_{\gamma_{\alpha,4}}}.
\end{align}
In the case of no correlation between the two estimators, $\rho_\alpha=0$ and the weight $\mu$ is equivalent to inverse variance weighting. If both estimators give an unbiased shear estimate, i.e., $\langle\widehat{\gamma}_{\alpha,2}\rangle = \langle\widehat{\gamma}_{\alpha,4}\rangle = \gamma_\alpha$ when averaged over an ensemble of galaxies, then the combined estimator also gives an unbiased shear estimate.

\section{Simulations and Tests}
\label{sec:simulation}
In this section, we describe mock astronomical images that we produce with known input shears to test the performance of our fourth-order shear estimator and compare it against the second-order shear estimator developed in \cite{fpfs}. We corrected for noise bias, selection bias, and detection bias using methods developed in Sec.~\ref{sec:method}. 
The estimated shear, $\hat{\gamma}_\alpha$, is related to the input shear, $\gamma_\alpha$, as
\begin{equation}
    \label{eq:sim:estimated_shear}
    \hat{\gamma}_\alpha = (1 + m_\alpha)\gamma_\alpha + c_\alpha,
\end{equation}
where the subscript $\alpha$ denotes each component of the shear and $m_\alpha$ (multiplicative bias) and $c_\alpha$ (additive bias) are used to test the accuracy of the shear estimator \citep{2006MNRAS.366..101H, 2006MNRAS.368.1323H}. We also combine second- and fourth-order shear estimators as derived in equation~\eqref{eq:shear_combined} and study the reduction in the shape noise.

\subsection{Simulation setup}
\label{sec:sim_setup}


We use the same simulation setup as used in \cite{fpfs}. In summary, we generate two sets of data: image simulations with isolated galaxies, where each realization contains multiple postage stamps that each have one galaxy, and blended galaxy image simulations, where galaxies are randomly distributed. 
We vary the seeing, pixel scale, and image noise level to match those of the HSC and LSST surveys. 

The pixel scale is set to $0\farcs168\,$ ($0\farcs2\,$) for HST (LSST) like simulation. The PSF for these simulations is modeled with a \cite{moffat} profile,
\begin{equation}
    p_m(\bm{x}) = \left[1 + c\left(\frac{|\bm{x}|}{r_p}\right)^2\right]^{-n},
\end{equation}
where $c$ and $r_p$ are set such that the full-width half maximum (FWHM) of the Moffat PSF is $0\farcs60$ ($0\farcs80$) matching the median seeing of actual HSC $i$-band (expected LSST $i$-band) images. The exponent $n$ is 3.5 for HSC simulations \citep{HSCY3ShapeCatalog} and 2.5 for LSST simulations \citep{metaDet_LSST2023}. Table~\ref{tab:simulation} summarizes the values used for each survey setup. To introduce an anisotropic PSF, we shear the PSF so that it has ellipticity $(e_1 = 0.02, e_2 = -0.02)$. We show a small region of one simulated image for both isolated and blended simulations in Fig.~\ref{fig:sim_img}.

\begin{table}
    \begin{tabular}{l c c}
        \hline
        Variable & HSC & LSST \\
        \hline
        Pixel scale & $0\farcs168$ & $0\farcs20$ \\
        Seeing & $0\farcs60$ & $0\farcs80$ \\
        Moffat profile exponent & $3.5$ & $2.5$\\
        Magnitude zero point & 27.0 mag & 30.0 mag\\
        Noise variance & $3.6\times10^{-2}$ & $3.5\times10^{-1}$\\
        PSF ellipticity & \multicolumn{2}{c}{$(e_1 = 0.02,~ e_2 = -0.02)$}
    \end{tabular}
    \caption{Table for simulation setups. The noise variance for each simulation is set to the mean noise variance in the three-year HSC data and the ten-year LSST observation, respectively. 
    }
    \label{tab:simulation}
\end{table}


When testing the accuracy of our fourth-order shear estimator, we use the same simulation used in \cite{fpfs}: for each type of simulation, we have two sets of galaxy images with the same realization of image noise generated; one distorted with $(\gamma_1 = 0.02, \gamma_2=0)$ and the other with $(\gamma_1 = -0.02, \gamma_2=0)$ as introduced in \citet{preciseSim_Pujol2019}. We used the ring test setup \citep{2007MNRAS.376...13M} by having our galaxy sample contain orthogonal galaxies with the same morphology and brightness but with the intrinsic major axes rotated by 90 degrees. This allows us to cancel out the intrinsic shape noise in our simulation.


\subsection{Isolated image simulations}
\label{sec:isolated_sim}
We test the performance of fourth order shear estimator relative to the second order estimator using isolated galaxy image simulations. The goal of this is to understand how the estimator performs when blending effects are absent, focusing on the impact of shear, PSF size, and image noise. For isolated galaxy image simulations, galaxy images are generated using the publicly available software \texttt{GalSim} \citep{galsim} and use the COSMOS HST parametric galaxy catalog \citep{mandelbaum_2019_3242143} with limiting magnitude $F814W = 25.2$ as the input galaxy catalog 
(see Section~3.1 of \cite{fpfs} for more details). Note that at this limiting magnitude for the input catalog, some of the lower SNR galaxies that are expected in the isolated simulations may be missing from the analysis. 
Galaxies are rendered in 64$\times$64~pix postage stamp images after the shear distortion and PSF convolution. Each realization (subfield) of the simulated image contains $100 \times 100$ postage stamps, and after including orthogonal galaxy pairs as described in Section~\ref{sec:sim_setup}, each subfield 
contains $5\times10^3$ orthogonal galaxy pairs. Each orthogonal galaxy is placed adjacent to its unrotated galaxy in the same simulation. We simulate 4000 subfields with different realizations for the noise image, the galaxy sample, and random galaxy rotation. 

\subsection{Blended image simulations}
\label{sec:lsst}
Blended galaxy simulations are essential for understanding the estimator's performance in realistic observational conditions, where the blending of multiple objects occurs. We also tested our fourth-order estimator on blended galaxy image simulations and compared it against the second-order estimator. For blended galaxy image simulations, we use the open-source package \texttt{descwl-shear-sims}\footnote{\url{https://github.com/LSSTDESC/descwl-shear-sims}} \citep{metaDet_LSST2023}, which includes the survey parameters for the simulation, such as noise and PSF. The package uses the input galaxy catalog from \texttt{WeakLensingDeblending}\footnote{\url{https://github.com/LSSTDESC/WeakLensingDeblending}}. 
The output of the package is a calibrated image with a subtracted background and estimated noise variance of the image. The simulated galaxies include bulge, disk, and AGN components. The bulge and the disk have varying fluxes and the AGN is represented as a point source located at the galaxy center. To include some of the complexity of realistic galaxy light profiles, the isophotes of these simulated galaxies are not strictly elliptical. Galaxy positions are randomly distributed across the image, without accounting for spatial clustering.


Each simulated image contains about $10^5$ input galaxies and covers 0.12 square degrees, corresponding to a galaxy number density of about 230 per square arcmin. However, not all of these galaxies are detectable at the depths achieved by HSC and LSST.  
When testing the accuracy of each shear estimator on blended galaxies, we also adopted the ring test to cancel out shape noise. For these blended simulations, the rotated galaxy is stored in a different realization. The blended galaxy image simulations are simulated with \texttt{descwl-shear-sims} \citep{metaDet_LSST2023}, and the setup is the same as \cite{fpfs3}.

For LSST-like simulations, we use $griz$ bands with the same galaxy profile and PSFs in each band without dithering. The noise variance level matches the expectations for a coadded image based on LSST ten-year observations, and each band has a different noise variance. The image is coadded in these four bands with inverse variance weighting based on the sky background noise. The multi-band coadds, then, have a well-defined PSF \citep{2023OJAp....6E...5M} since each band has a spatially constant weight across the images. These four bands were chosen as it is likely that LSST cosmic shear analysis will use observations in $griz$ bands. For the HSC-like setup, we used the image noise variance in the HSC $i$-band. HSC primarily uses the $i$-band for cosmic shear due to its better seeing conditions compared to other bands. 

\section{Results}
\label{sec:result}
In this section, we present the results of testing the precision and accuracy of our fourth-order \texttt{FPFS} shear estimator with isolated galaxies (Section~\ref{sec:result:isolatedG}) and blended galaxies (Section~\ref{sec:result:bg}) as described in Section~\ref{sec:simulation}.  We also compare the results against the second-order \texttt{FPFS} shear estimator and combine the two shear estimators to quantify the reduction in shape noise.

\begin{figure}
    \centering
    \includegraphics[width=0.5\textwidth]{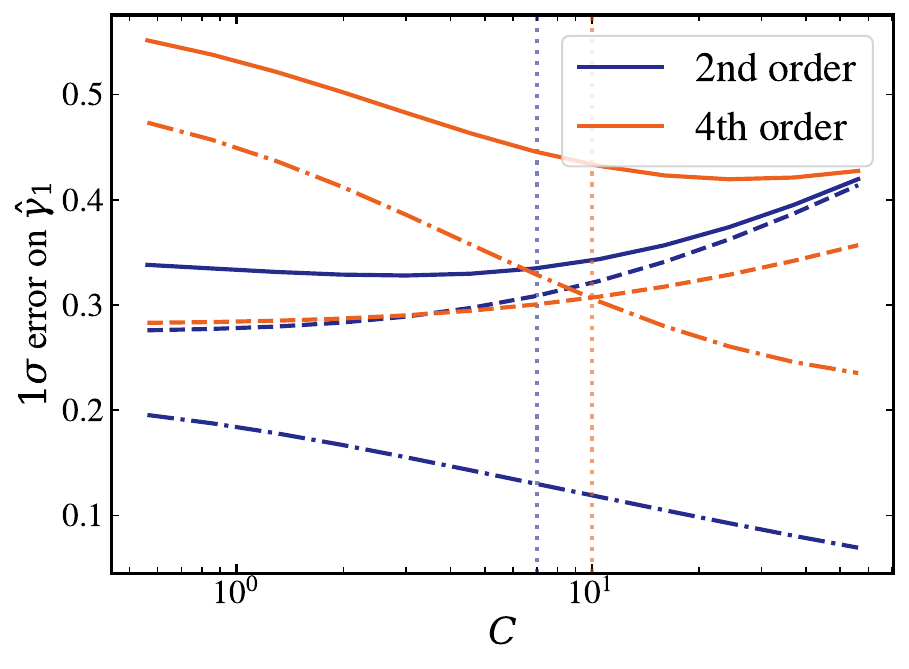}
    \caption{The $1\sigma$ statistical uncertainty on a single component of the estimated shear $\hat{\gamma}_1$ for individual isolated galaxies (solid lines) as a function of the weighting parameter, $C^{(n)}$, in equation~\eqref{eq:ellipticity_fourth}. 
    For each second (blue) and fourth (orange) order estimator, the total uncertainty has contributions due to image noise (dotted lines) and intrinsic shape noise (dash-dotted lines). The vertical dashed lines show the values of $C^{(n)}$ used for each estimator throughout the paper. }
    \label{fig:fpfsvsc}
\end{figure}

\subsection{Isolated Galaxies}
\label{sec:result:isolatedG}
In this subsection, we focus on isolated galaxies simulated within postage stamps as described in Section~\ref{sec:isolated_sim}. 

\subsubsection{Precision Test}
\label{sec:results:isolatedG_precision}
We first tested the performance of our fourth-order estimator and compared the uncertainty on this new estimator to that of the second-order estimator.  We then computed how much information is gained, or how much uncertainty is reduced, by combining the two estimators.
When analyzing the uncertainty of each estimator, we use 100 out of 4000 subfields to save computational time.  
For each subfield we include the intrinsic shape noise in our analysis by excluding a single randomly chosen galaxy from each orthogonal pair when both are detected. 
In addition, we deliberately positioned the center of each galaxy at the center of the postage stamp. We forced the pipeline to do measurements based on the known center without running any detection and selection process during the image processing. This is equivalent to setting the selection weight function, $w$ in equation \eqref{eq:selection_weight}, to $1$ for all galaxies. We use the results from this process to set the \texttt{FPFS} weighting parameter $C^{(n)}$ 
in the denominator of the spin-2 ellipticity estimator in equation~\eqref{eq:ellipticity_fourth}.

\begin{figure}
    \centering
    \includegraphics[width=0.5\textwidth]{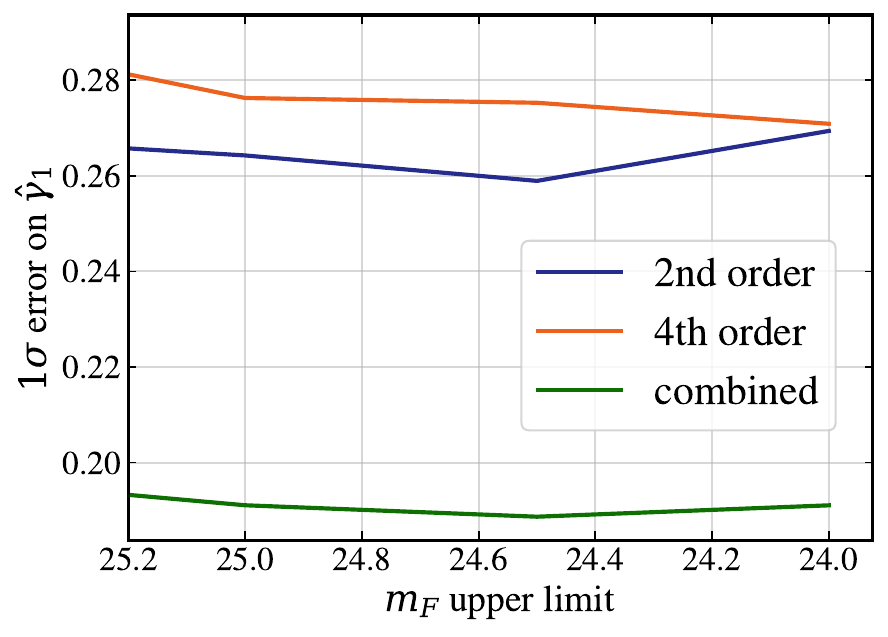}
    \caption{The $1\sigma$ statistical uncertainty on shear measurement $\hat{\gamma}_1$ for individual isolated galaxies as a function of the upper limit of \texttt{FPFS} magnitude ($m_F$) for the galaxies included in the measurement.  
    The uncertainty for each second (blue) and fourth (orange) order estimator includes contributions due to image noise and intrinsic shape noise. Equation~\eqref{eq:shear_combined} is used to combine the two shear measurements (green). }
    \label{fig:isolatedG_prec}
\end{figure}
For each subfield $i$, the estimated shear is computed as:
\begin{equation}
    \label{eq:measured_g}
    \hat{\gamma}^{(n)}_{\alpha,i} = \frac{\sum_{j\in i} \tilde{\tilde{w}}^{(n)}_{j} \:\tilde{\tilde{{e}}}^{(n)}_{\alpha,j}}{\sum_{j\in i} \tilde{\tilde{R}}^{(n)}_{\alpha,j}},
\end{equation}
where $n\in\{2,4\}$ is the order of the shear estimator, $\alpha$ is the component of the shear, and the summation is over all galaxies $j$ in subfield $i$. Note that the weight is also order-dependent as the weight parameter $C^{(n)}$ is different for each order. Forcing the measurements at the known centers and not applying any selection criteria is equivalent to setting the effective weight in the shear estimator $\tilde{\tilde{w}}^{(n)}_{j} = 1$. 

We calculate the statistical uncertainty in the shear estimation from a population of subfields of galaxies as:
\begin{equation}
\label{eq:uncertainty}
    \left(\sigma_\gamma^{(n)}\right)^2 = \left<\frac{1}{2}\left(\left(\hat{\gamma}^{(n)}_1\right)^2+\left(\hat{\gamma}^{(n)}_2\right)^2\right)\right> N,
\end{equation}
where $N$ is the number of galaxies in each subfield. With each subfield having $100\times100$ postage stamps, but with one galaxy in each orthogonal pair excluded, $N = 5 \times 10^3$. 
The total uncertainty is a combination of uncertainty due to galaxies' intrinsic shape noise, $\gamma^{(n)}_{\texttt{RMS}}$, and uncertainty due to noise in the galaxy images, $\sigma_e^{(n)}$. We assume that they add in quadrature, such that
\begin{equation}
    \label{eq:uncertainty_num}
     \left(\sigma_\gamma^{(n)}\right)^2 = \left(\gamma^{(n)}_{\texttt{RMS}}\right)^2 + \left(\sigma_e^{(n)}\right)^2\,.
\end{equation}

To obtain the galaxies' intrinsic shape noise, $\gamma^{(n)}_{\texttt{RMS}}$, we use simulated galaxies with zero image noise and the same weighing parameter value $C^{(n)}$ for each estimator to measure the \texttt{FPFS} ellipticity. 
To save computational time, we only measure the first component of the ellipticity and calculate the RMS across our sample of subfields. To obtain the measurement uncertainty, $\sigma_e^{(n)}$, we measure the total uncertainty using a noisy realization of the galaxy image, again using only the first component of the ellipticity. We then subtract $e_{\texttt{RMS}}$ in quadrature following equation~\eqref{eq:uncertainty_num} to get the measurement uncertainty for each shear estimator.


The statistical uncertainty 
of \texttt{FPFS} also depends on the \texttt{FPFS} weighting parameter, $C^{(n)}$, in equation~\eqref{eq:ellipticity_fourth} 
\citep{fpfs}. Here, we assume image noise fields that are homogeneous but that have pixel-to-pixel correlations. This is a reasonable approximation for faint, small galaxies observed in ground-based surveys, where noise is primarily due to sky background fluctuations and where measurements may be made on a coadd with pixel-to-pixel noise correlations. However, this assumption may not fully capture the noise properties in upcoming Stage IV imaging surveys. Further testing is required to evaluate the robustness of this assumption to achieve the precision needed for percent level cosmology. 
In Fig.~\ref{fig:fpfsvsc}, we show each source of uncertainty as a function of $C^{(n)}$ for the second-order (blue) and fourth-order (orange) estimators on simulated isolated galaxies. 
 The galaxies' intrinsic shape noise (dashed) for both estimators increases as a function of $C^{(n)}$, whereas the measurement uncertainty (dashed-dotted) for both decreases as a function of $C^{(n)}$ since a larger value of $C^{(n)}$ adds more weight to the limited number of bright galaxies with higher SNR. The measurement uncertainty for the fourth-order estimator is larger than that of the second-order for our simulation setup as the fourth-order shapelet basis amplifies the noise. 
 From this result, we set $C^{(2)} = 7$ for the second-order shear estimator and $C^{(4)} = 10$ for the fourth-order shear estimator for the rest of the paper.

Once we fixed the value of $C^{(n)}$ for each shear estimator, we relaxed the forced measurement and allowed the pipeline to carry out detection and selection during the image processing. For the rest of this section, during the image production step and before running the \texttt{FPFS} detection process, we shift the galaxy centroid with random sub-pixel offsets. 
Following \citet{fpfs}, we run the \texttt{FPFS} peak detection algorithm and use the detected peaks as the centroid. Using the detected peaks to measure the galaxy properties, we then apply flux- and resolution-based selection criteria in equation~\eqref{eq:selection_weight} to assign selection weights to the detected galaxy sample. After analytically correcting for noise bias, selection bias, and detection bias, we show the shape measurement uncertainty from the second-order (blue) and fourth-order (orange) estimator on isolated galaxies in Fig.~\ref{fig:isolatedG_prec}. This figure shows that for isolated galaxy samples, combining the two shear estimators reduces the statistical uncertainty in shear inference by $\sim$30\%. 
This result suggests that for isolated galaxies, 
we can get significantly more information if we estimate the shear using second and fourth-order moments and combine them using equation~\eqref{eq:shear_combined_error}.  
A $30$ per cent reduction in shape noise is equivalent to the increase in sample size that would be achieved by expanding the survey area by $70$ per cent.


\subsubsection{Adaptive vs.\ Non-adaptive moments}
\label{sec:result:isolatedG_regauss}


We next compare each \texttt{FPFS} shear estimator against the widely used \texttt{reGauss} shear estimator method \citep{regauss}. This comparison is motivated by the differences in how these methods use moments of the galaxy profiles to measure ellipticity. One key distinction between the two is the use of adaptive vs.\ non-adaptive approaches. In the non-adaptive method, as implemented in \texttt{FPFS}, a fixed kernel (or weight function) is applied uniformly across all galaxies, regardless of their morphology. The adaptive method used by \texttt{reGauss} optimizes the weight function for each galaxy based on its specific shape and size. This optimization improves the precision of the shear estimation by effectively accounting for variations in galaxy morphology to match the weight functions to the galaxy images. 
We use the \textsc{GalSim} implementation of \texttt{reGauss} to correct for the PSF when estimating galaxy shapes. An important output of the \texttt{reGauss} estimator is the components of the spin-2 ellipticity of each galaxy:
\begin{equation}
    (e_1, e_2) = \frac{1 - (b/a)^2}{1 + (b/a)^2}(\cos 2\phi, \sin2\phi),
\end{equation}
where $b/a$ is the axis ratio and $\phi$ is the position angle of the major axis with respect to sky coordinates.
Another important output of \texttt{reGauss} is the resolution factor $R_2$:
\begin{equation}
    R_2 = 1 - \frac{T_\mathrm{PSF}}{T_\mathrm{gal}},
\end{equation}
where $T_\mathrm{PSF}$ is the trace of the second moment matrix of the PSF and $T_\mathrm{gal}$ is the trace of the second moment matrix of the observed PSF-convolved galaxy image. The resolution factor quantifies how well the galaxy image is resolved relative to the PSF; $R_2\sim1$ for well-resolved objects and $R_2\sim0$ for poorly resolved images. In general, the positively and negatively sheared galaxy images have different resolution factors $(R^+_2 \neq R^-_2)$. 

We select well-resolved galaxies to avoid a large multiplicative bias correction at a low resolution factor for \texttt{reGauss} \citep{HSCY3ShapeCatalog}. 
To ensure a consistent sample between \texttt{FPFS} and \texttt{reGauss}, we apply the following criteria for each subfield: only keep galaxies that satisfy $R_2^{(+,-)} > 0.3$ and $|e^{(+,-)}| < 2$. 
Applying the resolution factor and the ellipticity criterion reduces the sample to $\sim$$4\times 10^{3}$ galaxies 
out of $5\times10^3$ galaxies in each subfield. When comparing the two shear estimation methods, we only retain galaxies that were detected in both methods, ensuring the same sample population for both \texttt{FPFS} and \texttt{reGauss}. This helps isolate differences in precision between the methods without introducing sample bias. 

For an isotropically oriented galaxy ensemble distorted by some known constant shear, the shear responsivity can be estimated:
\begin{equation}
    \label{eq:shear_response_regauss}
    \hat{R}^{(R)} = \frac{\left\langle \widehat{w^{(R)} e_1^{(R)}}^+ - \widehat{w^{(R)} e_1^{(R)}}^- \right\rangle}{2\gamma},
\end{equation}
where the superscript ``+'' (``-'') refers to measurements of galaxies with positive (negative) shear $\gamma$ applied, $R$ denotes quantities estimated using \texttt{reGauss} method, and $w^{(R)}$ is the inverse variance weight defined as
 \begin{equation}
    \label{eq:weight_regauss}
    w^{(R)} = \frac{1}{\left(\gamma^{(R)}_\texttt{RMS}\right)^2 + \left(\sigma_\gamma^{(R)}\right)^2},
\end{equation}
Using the \texttt{reGauss} method, the shear is then estimated as
\begin{equation}
    \label{eq:measured_g_regauss}
    \hat{\gamma}^{(R)}_{\alpha,i} = \frac{\sum_{j\in i} {w^{(R)}_j} \:{e}^{(R)}_{\alpha,j}}{R^{(R)}_{\alpha,j}},
\end{equation}
To obtain $\gamma^{(R)}_{\texttt{RMS}}$, we use the noise-free isolated galaxy image simulation and set the weights $w = 1$. We use equation~\eqref{eq:measured_g_regauss} to estimate the shear and take the variance across multiple subfields to get the uncertainty. This process gives $\gamma^{(R)}_{\texttt{RMS}}=0.26$, the typically used value for intrinsic shape noise. To obtain the total uncertainty including intrinsic shape noise and measurement error, we use noisy realizations of the images. We set $\gamma^{(R)}_{\texttt{RMS}} = 0.26$ and use the value $\sigma_\gamma^{(R)}$ returned from \texttt{GalSim}.

\begin{figure}
    \centering
    \includegraphics[width=0.5\textwidth]{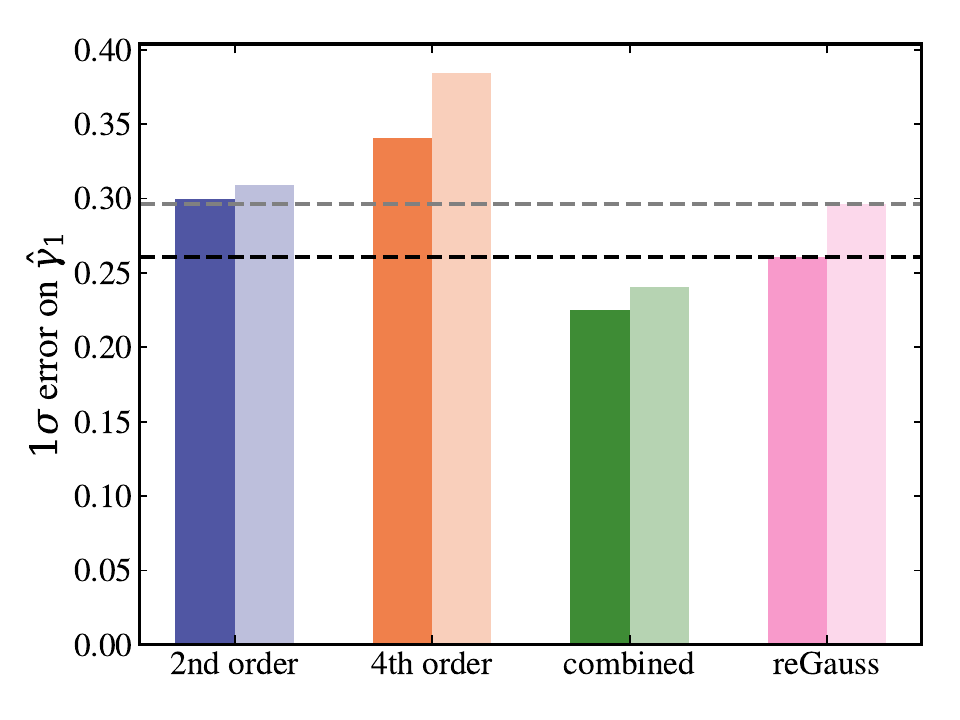}
    \caption{The $1\sigma$ statistical uncertainty on shear measurement of individual galaxies from each estimator; second (blue) and fourth (orange) order, combining the two \texttt{FPFS} estimators (green), and from \texttt{reGauss} (pink) method. The uncertainty values of \texttt{reGauss} are drawn in dotted horizontal lines to simplify comparison. For each estimator, the left bar (deep) only has contributions from the intrinsic shape noise and the right bar (shallow) has contributions from intrinsic shape noise and image noise. }
    \label{fig:regauss}
\end{figure}

In Fig.~\ref{fig:regauss}, we compare the $1\sigma$ statistical uncertainty on the shear measurement of individual galaxies from different shear estimation methods for 100 subfields. 
For each shear estimation method, we present the intrinsic shape noise (shaded) and the total uncertainty (no shade), which includes intrinsic shape noise and measurement error. We use $C^{(2)} = 7$ and $C^{(4)} = 10$ 
for the second order and fourth order \texttt{AnaCal} estimator, respectively, which optimizes the total uncertainty in Fig.~\ref{fig:fpfsvsc}. Equation \eqref{eq:shear_combined} is used to combine the two \texttt{AnaCal} estimators and equation \eqref{eq:shear_combined_error} to quantify the uncertainty of the combined estimator. 
The dashed horizontal lines are drawn to indicate the values for \texttt{reGauss}. In general, \texttt{reGauss} performs better than individual \texttt{FPFS} estimators, or in other words the adaptive moments method has lower uncertainty than the non-adaptive moments method. However, when combining the second and fourth-order \texttt{FPFS} estimators, we get $\sim30\%$  reduction in the shape noise compared to that of \texttt{reGauss}.


\begin{figure*}
    \centering
    \includegraphics[width=\textwidth]{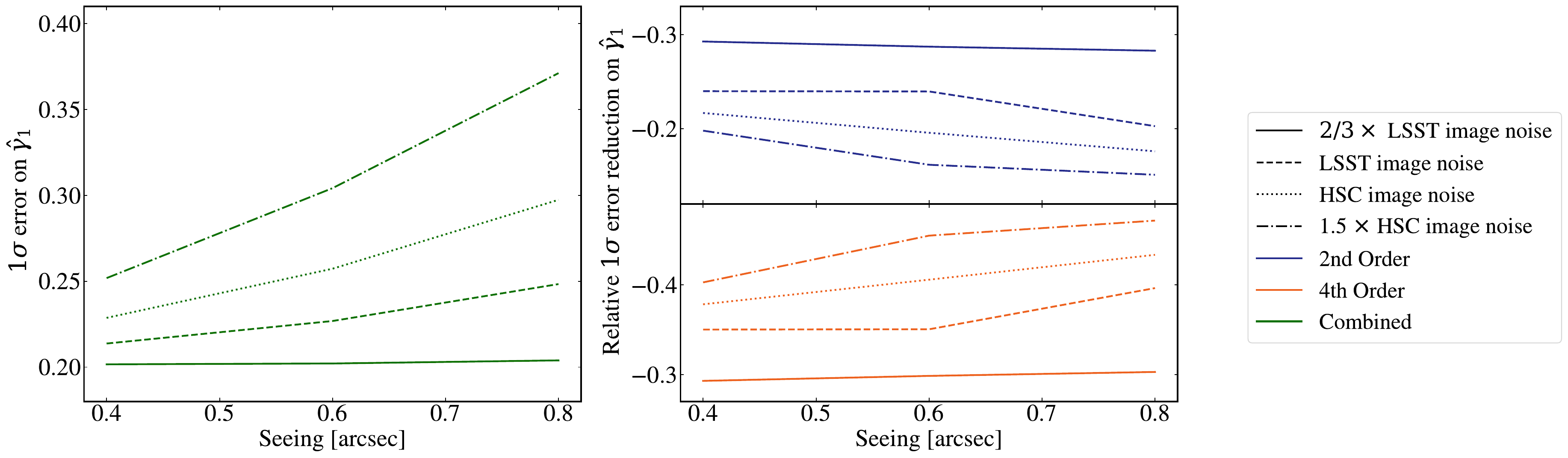}
    \caption{The $1\sigma$ statistical uncertainty per galaxy on shear measurement $\hat{\gamma}_1$ for isolated galaxies as a function of the PSF size using different image noise levels and shear estimators. The uncertainty for each second (blue) and fourth (orange) order estimator includes contributions due to image noise and intrinsic shape noise. Equation~\eqref{eq:shear_combined} is used to combine the two shear measurements (green). Left panel shows the $1\sigma$ error on $\hat{\gamma}_1$ using the combined shear estimator. Right panel shows the ratio of statistical uncertainty from the second order and fourth order estimators compared to the combined estimator. The noise variance for HSC is approximately twice that of the LSST noise variance. The results indicate that the combined shear estimator consistently provides lower statistical uncertainty relative to individual shear estimators across varying seeing size and image noise levels. 
 }
    \label{fig:iso_seeing_noise}
\end{figure*}

\subsubsection{Varying Seeing and Image Noise}
\label{sec:result:varyingseeing}
In the previous subsections, we used the HSC PSF size ($0\farcs60$) and the average noise in the $i$-band for isolated image simulations, and found a $\sim30\%$ reduction in shape noise when including the 4th moment-based shear estimator along with the second moment-based estimator. In this subsection, we vary the PSF size and image noise to see test the dependence of this finding on the observational conditions. In Fig.~\ref{fig:iso_seeing_noise}, we show the uncertainty in each shear estimator and combined for four different PSF seeing sizes and image noise levels.  
``$1.5\: \times $ HSC'' corresponds to an image noise level with a noise variance that is 1.5 times the average noise variance of HSC in $i-$band coadds,  
and ``$2/3\: \times $ LSST'' corresponds to the noise variance that is $2/3$ of the expected noise variance from ten years of LSST coadds. The reduction in shape noise is consistent for all different seeing sizes and image noise levels.  The curve for ``$2/3 \times$ LSST'' is basically flat because for a low image noise level, we can perfectly deconvolve the image with a PSF without amplifying noise on small scales. 
As illustrated in the right panel of Fig.~\ref{fig:iso_seeing_noise}, the ratio of uncertainty from second/fourth order shear estimators to that of combined is roughly independent of PSF size and image noise level, and the fourth order estimator exhibits a greater susceptibility to image noise than the second order estimator. This suggests that for isolated galaxies, combining the two independent \texttt{FPFS} shear estimators consistently reduces statistical uncertainty, making it the preferred approach. 

\subsubsection{Accuracy Test}
\label{sec:result:isolatedG_acc}

\begin{figure*}
    \centering
    \includegraphics[width=0.8\textwidth]{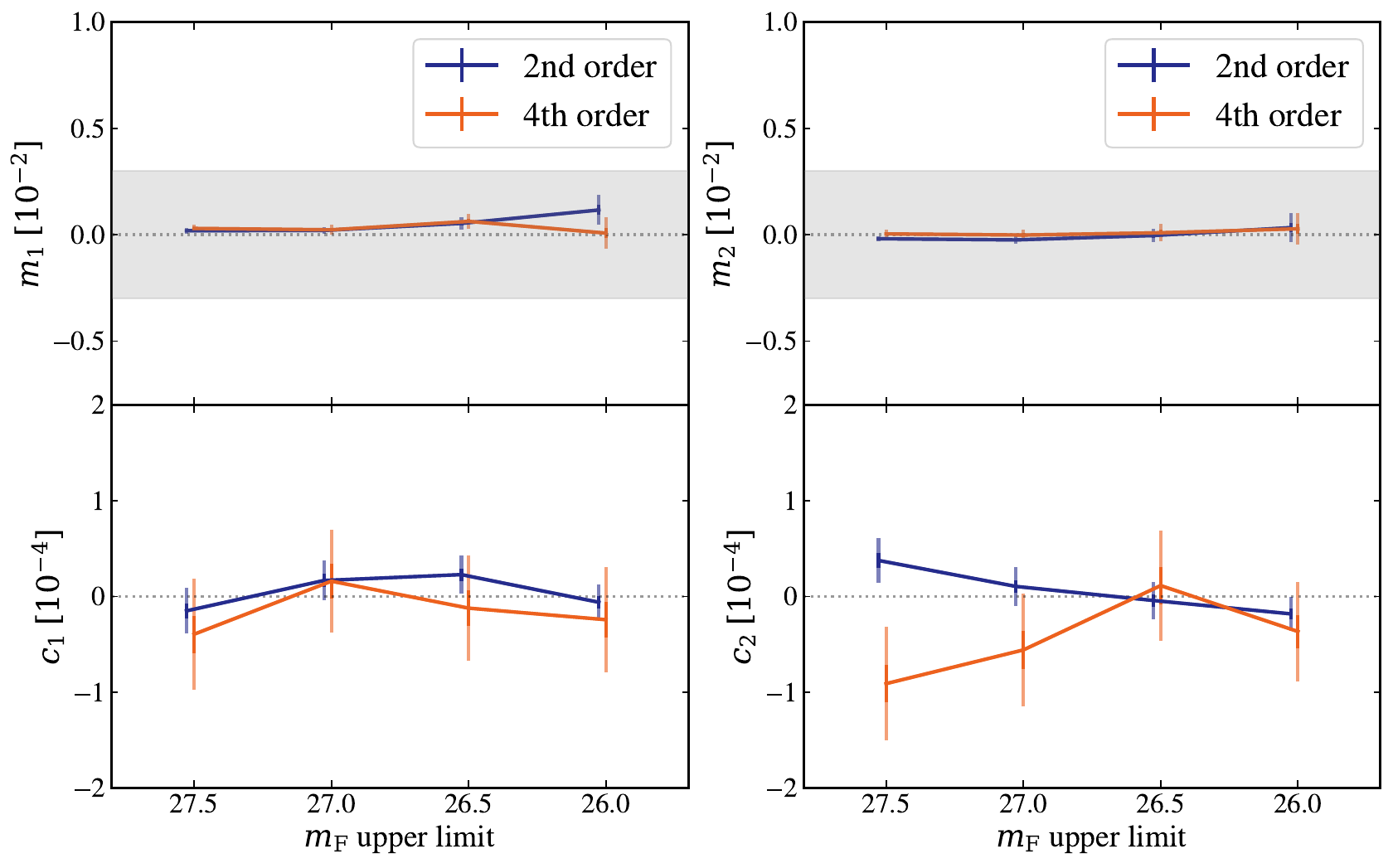}
    \caption{The multiplicative bias (upper panel) and additive bias (lower panel) of the \texttt{AnaCal} shear estimator on isolated galaxies using an HSC-like configuration. 
    Blue (orange) lines are results using the second (fourth) order shear estimator. The error bars show the $1\sigma$ and $3\sigma$ uncertainties, respectively, and the gray-shaded region is the LSST ten-year requirement on the multiplicative shear bias \citep{2018arXiv180901669T}. In both panels, the blue points are slightly shifted to differentiate the error bars. Both estimators give unbiased shear estimation for each shear component. 
    }
    \label{fig:isolated_g1g2}
\end{figure*}
In this subsection, we test how accurate our fourth-order estimator is compared to the second-order estimator. The multiplicative and additive biases in our shear estimation in equation~\eqref{eq:sim:estimated_shear} are measured as introduced in \citet{metaDet_Sheldon2020} as
\begin{equation}
\label{eq:mbias_estimator}
    {m_\alpha} = \frac{\langle \widehat{w e_\alpha}^+ - \widehat{w e_\alpha}^- \rangle}
    {0.02\left\langle \widehat{R_{\alpha}}^{+} + \widehat{R_{\alpha}}^{-} \right\rangle}
    -1\,,
\end{equation}
and
\begin{equation}
\label{eq:cbias_estimator}
    {c_\alpha} = \frac{\langle \widehat{w e_\alpha}^+ + \widehat{w e_\alpha}^- \rangle}
    {\left\langle \widehat{R_{\alpha}}^{+} + \widehat{R_{\alpha}}^{-} \right\rangle},
\end{equation}
where $0.02$ is the value of constant shear applied to each galaxy image and $\widehat{w e_1}$ and $ \widehat{R_{1}}$ are the first components of the weighted ellipticity and its shear responsivity after noise bias correction, respectively. When quantifying the multiplicative and additive biases, we test $\gamma_1$ and $\gamma_2$ separately; when testing $\gamma_1$ we set $\gamma_2 = 0$ and vice versa. The quantities with superscript ``$+$'' are estimated from images distorted by positive shear, $(\gamma_1 = 0.02$ and $\gamma_2 = 0)$ or $(\gamma_1 = 0$ and $\gamma_2 = 0.02)$ and superscript ``$-$'' are estimated from images distorted by negative shear, $(\gamma_1 = -0.02$ and $\gamma_2 =0)$ or $(\gamma_1 = 0$ and $\gamma_2 = -0.02)$. It is worth noting that galaxies in each orthogonal galaxy pair and galaxies with different applied shears (positive and negative $\gamma_1$) are selected and weighted independently. We then apply the shear estimator to this selected sample to evaluate our corrections for detection and selection biases. 

For the accuracy test, we used the HSC seeing size and image noise level and used 4000 subfields with galaxy orthogonal pairs to approximately cancel out the intrinsic shape noise. 
The amplitudes of the multiplicative (additive) biases for each \texttt{FPFS} estimator are shown in the top (bottom) panel of Fig.~\ref{fig:isolated_g1g2}. We found that the multiplicative biases for both estimators are well within the LSST ten-year requirements, and we do not find a significant additive bias. 

\subsection{Blended Galaxies}
\label{sec:result:bg}
In this subsection, we discuss the results of the precision test (as shown for isolated galaxies in Sec.~\ref{sec:result:bg_prec}) and the accuracy test (Sec.~\ref{sec:result:bg_acc}) by applying the second and fourth-order  \texttt{AnaCal} shear estimators on the blended simulated galaxy images as described in Sec.~\ref{sec:lsst}. In \cite{fpfs3}, they demonstrated that the second order \texttt{AnaCal} shear estimator 
yields unbiased results for the accuracy test. The results presented here confirm that the second order shear estimator tested in this section is consistent with these previous performance tests of the \texttt{AnaCal} second order shear estimator. 

\subsubsection{Precision Test}
\label{sec:result:bg_prec}
\begin{figure*}
    \centering
    \includegraphics[width=0.8\textwidth]{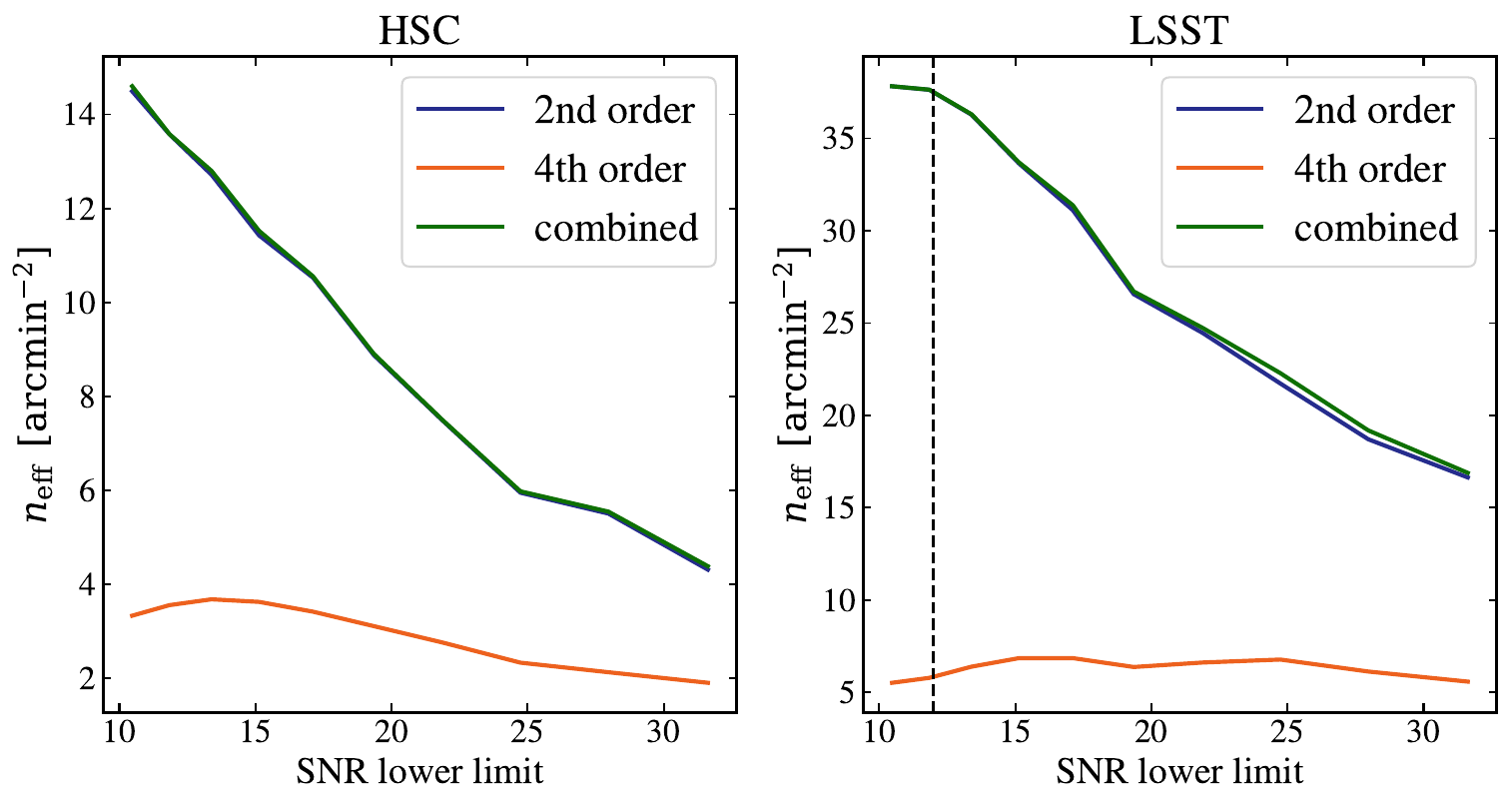}
    \caption{The effective galaxy number density as a function of SNR cut for each \texttt{AnaCal} estimator and for each simulation setup (PSF seeing size and image noise level) using 100 subfields with blended galaxies. The statistical uncertainties that go into this effective number density include intrinsic shape noise, image noise, and noise from blending.  
    The vertical dashed line represents the default SNR cut used in this paper. 
    }
    \label{fig:neff}
\end{figure*}

When quantifying the uncertainty on each estimator for noisy simulated blended galaxy images, we use 100 subfields. We measure shear from each image using both the second and fourth-order \texttt{FPFS} shear estimators and use the scatter in these 100 shear measurements to derive the statistical error. We obtain the per-component statistical uncertainty $\sigma_\gamma$ as defined in equation~\eqref{eq:uncertainty} on shear estimation for a region of one square arcmin by normalizing the statistical error according to the area of each simulated image. We follow \citet{fpfs3} to derive the $n_\text{eff}$ from blended galaxy image simulations as

\begin{equation}
    \label{eq:neff}
    n_\text{eff} = \left(\frac{0.26}{\sigma_\gamma}\right)^2 [\text{arcmin}^{-2}],
\end{equation}
where 0.26 is the per component root-mean-square (RMS) of intrinsic shape noise using the widely used \texttt{reGauss} shear estimation method (see Fig.~\ref{fig:regauss}). 
We choose to use 0.26 for the shape noise RMS so that we can compare our estimation of the effective number density with other methods that use \texttt{reGauss}  
Typically, the galaxy number density is estimated by counting the galaxy number (with weights) after detecting galaxies from images \citep{WLsurvey_neffective_Chang2013, 2018arXiv180901669T}. Doing so assumes that the statistical uncertainty in shear estimation for each galaxy is uncorrelated. Effectively we compute the equivalent galaxy number density for the shear if it were to be estimated with \texttt{reGauss}.

Motivated by the result of Fig.~\ref{fig:iso_seeing_noise}, we show the $n_\mathrm{eff}$ as a function of SNR cut for each shear estimator for two different survey setups, HSC and LSST, in Fig.~\ref{fig:neff}. Each survey setup follows the respective PSF size and image noise level. 
The $n_\mathrm{eff}\sim15\: \mathrm{arcmin}^{-2}$ for HSC setup is $35\%$ smaller than the $n_\mathrm{eff} \sim 20 \:\mathrm{arcmin}^{-2}$ for a PSF size of 0.6~arcsec from HSC-Y3 \citep{HSCY3ShapeCatalog}.  
The value may be smaller because we did not optimize the smoothness parameter for detection. 
Optimizing this parameter will improve the $n_\mathrm{eff}.$ The value of $n_\mathrm{eff}\sim39\: \mathrm{arcmin}^{-2}$ for LSST seeing size and image noise level is $5\%$ higher than $n_\mathrm{eff}$ estimated from \cite{WLsurvey_neffective_Chang2013}. They reported $n_\mathrm{neff}\sim37 \:\mathrm{arcmin}^{-2}$ for the expected distributions of observing parameters and all lensing data ($r$ and $i$ band) before considering blending and masking.  
The value is estimated in this work is slightly higher due to the use of different selection cuts and different numbers of bands to compute the value; \cite{WLsurvey_neffective_Chang2013} combined $r$- and $i$-bands, whereas we used \textit{griz}-band coadded images.


We find that for the fourth-order \texttt{FPFS} shear estimator, the effective number density is relatively constant for every SNR lower limit, as the fourth-order amplifies measurement error caused by image noise.
This shows that the fourth-order \texttt{FPFS} shear estimator is more sensitive to blending. By combining second- and fourth-order, we find that there is $\sim 2\%$ improvement in the effective number density. 
When estimating shear using blended galaxy samples, the fourth order shear estimator adds minimal information to the second-order estimator. The sensitivity of the fourth order estimator to noise and blending highlights its limitations in current ground-based surveys. Space-based images, and deep-field images with higher SNR, offer the potential for improving the effective number density using the second- and fourth-order shear estimators together. We defer this work to future studies.


\subsubsection{Accuracy Test}
\begin{figure}
    \centering
    \includegraphics[width=0.5\textwidth]{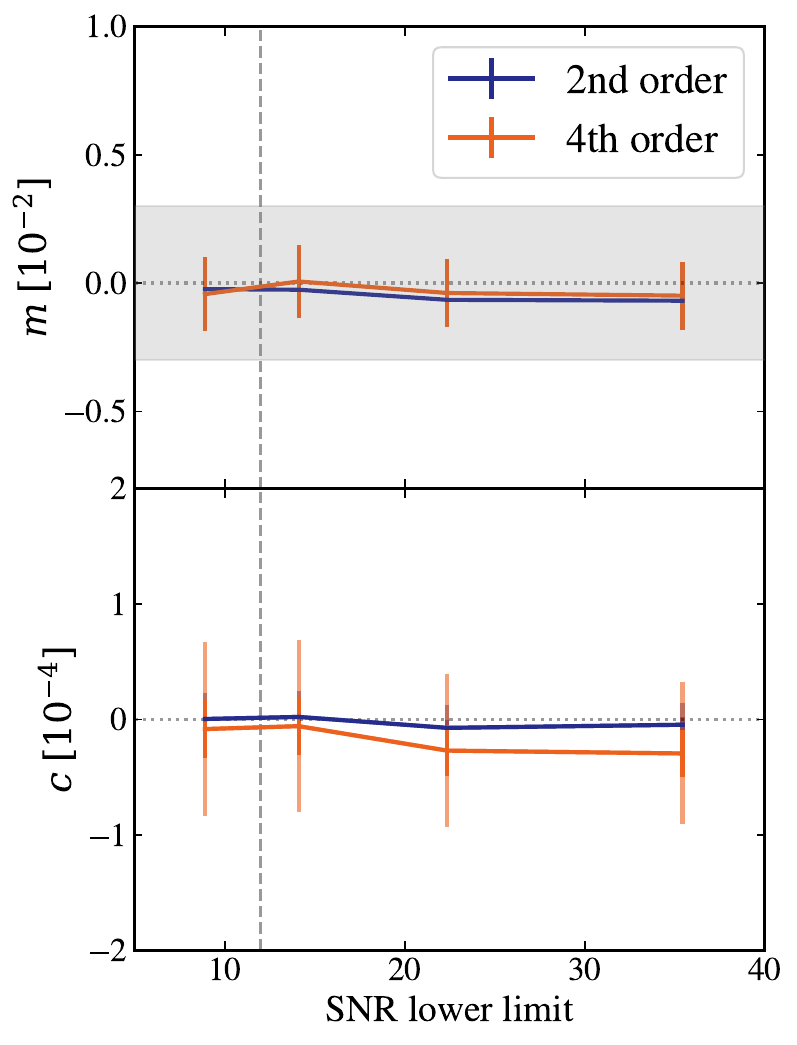}
    \caption{The multiplicative bias (upper panel) and additive bias (lower panel) of the \texttt{AnaCal} shear estimator on blended galaxies using an LSST-like configuration. 
    Blue (orange) lines show the second (fourth) order shear estimator. The error bars show the $1\sigma$ and $3\sigma$ uncertainties, respectively, and the shaded gray region is the LSST ten-year requirement on the shear multiplicative bias. The vertical dashed line indicates our default selection SNR $ > 12$.
}
\label{fig:blended_acc}
\end{figure}

We also tested the accuracy of the fourth-order shear estimator on blended galaxies. We used equations~\eqref{eq:mbias_estimator} and~\eqref{eq:cbias_estimator} to measure multiplicative and additive biases in our shear estimator. We include the orthogonal galaxy pairs to approximately cancel out the intrinsic shape noise and use all 5000 subfields to tightly constrain the shear biases. 
To save computational time, we only test the noise bias correction for the estimation of $\widehat{\gamma}_1$ and remove the subscript ($m \equiv m_1$ and $c \equiv c_1$); the multiplicative and additive biases of $\widehat{\gamma}_2$ should be comparable to those of $\widehat{\gamma}_1$. Fig.~\ref{fig:blended_acc} shows the multiplicative and additive biases for each \texttt{AnaCal} estimator with the LSST ten-year requirement. The result shows that the fourth-order shear estimator satisfies the LSST ten-year requirement even in the presence of blending. 


\label{sec:result:bg_acc}



\section{Conclusion}
\label{sec:conclusion}

In this work, we extended the perturbation-based shear estimator \texttt{FPFS}, which is part of the new suite of analytic shear estimation algorithms \texttt{AnaCal}, and developed a higher order \texttt{FPFS} shear estimator.  We  compared its performance against the second order shear estimator developed in \cite{fpfs}. The \texttt{AnaCal} framework analytically corrects for the detection and selection bias in shear estimation by deriving the shear response of image pixels. It also uses the \textit{renoising} approach \citep{fpfs4} to correct for the noise bias by adding an additional layer of noise to the image. This method does not require any computation of derivatives of noisy nonlinear observables. This work focuses on evaluating the accuracy and precision of our new shear estimator on constant-shear simulations, in which all galaxies in isolated and blended image simulations undergo the same shear. The application of this work with redshift-dependent shear will be explored in future research.

Typically, the shear estimation uses the lowest order (specifically, second order) to define spin-2 observables. We used fourth-order \texttt{FPFS} shapelet moments to define our spin-2 observables (e.g., ellipticity) and incorporated the fourth-order shear estimator into the \texttt{FPFS} framework. 
We tested the precision and accuracy of our new shear estimator with HSC-like and LSST-like simulations on both isolated and blended galaxy image simulations. We find that for both sets of image simulations, the multiplicative shear bias $|m|$ for the fourth-order shear estimator is consistently less than $3\times10^{-3}$ (LSST ten-year requirement on the control of multiplicative bias) within the $3\sigma$ uncertainties.

When the fourth-order shear estimator is combined with the second-order shear estimator, we found that the shape noise is reduced by $\sim35\%$ for isolated galaxy image simulations compared to using the second-order estimator alone.  For isolated galaxies, the fourth-order shear estimator detects information that is not probed by the second moment, and we found that the shear estimators are complementary. We also found that the effective number density is improved by $\sim2\%$ for blended galaxy image simulation. When blending is introduced, the fourth order estimator treats the blended galaxies as noise and amplifies their signals causing them to behave like noise. 
We recommend combining both the second- and fourth-order \texttt{FPFS} shear estimators when estimating shear for high SNR and isolated galaxy ensembles. While this approach shows promise, further testing is required for its application to deblended samples and is left for future work.  

An important target for future work is testing this method on spaced-based image simulations, as there may be less blending of the galaxies. In that regime, it would be worth evaluating the improvement in $n_{\text{eff}}$ from combining the two orders of shear estimators. This paper focused on evaluating and testing the new shear estimator using constant-shear simulations, in which all isolated and blended galaxies within a single image were subject to uniform shear distortions. \cite{MacCrann_redshiftdependentshear} showed and tested the performance of \textsc{Metacalibration} using simulations with redshift-dependent shear by dividing galaxies into four redshift bins and applying different shear values to each bin. Their findings indicate that the blending of galaxies at different redshifts changes the effective number density distribution as a function of redshift. The implications of redshift-dependent shear within this framework are left for future work. 

Another potential extension of this framework involves integrating it with the approach outlined in \cite{Tianqing_GeneralFrameworkPSF}, which can detect PSF leakage and modeling errors from all spin-2 quantities contributed by the PSF second and higher-order moments. Combining the second and fourth order shear estimators with this framework should give more information on  the impact of PSF systematics on cosmic shear analyses and how to mitigate their impact on cosmological analysis.  A detailed exploration is deferred to future work.

\section*{Data Availability}

No new data was generated for this work. The code used to analyze simulated galaxy images and obtain shape catalogs is available from \url{https://github.com/mr-superonion/xlens/tree/v0.2.2} and the code for shear estimation is available from \url{https://github.com/mr-superonion/AnaCal/tree/v0.2.2}.

\section*{Acknowledgement}
This work was performed using the Vera cluster at the McWilliams Center for Cosmology at Carnegie Mellon University, operated by the Pittsburgh Supercomputing Center facility. The authors would like to thank Scott Dodelson, Gary Bernstein, and Alexandre Refregier for their useful discussions.

We thank the maintainers of numpy \citep{harris2020array}, \textsc{SciPy} \citep{2020SciPy-NMeth}, Matplotlib \citep{Hunter:2007}, and \textsc{GalSim} \citep{galsim} for their excellent open-source software.

Andy Park, Xiangchong Li, and Rachel Mandelbaum are supported in part by the Department of Energy grant DE-SC0010118 and in part by a grant from the Simons Foundation (Simons Investigator in Astrophysics, Award ID 620789). Xiangchong Li is an employee of Brookhaven Science Associates, LLC under
Contract No. DE-SC0012704 with the U.S. Department of Energy.



\bibliographystyle{mnras}
\bibliography{example} 

\begin{thebibliography}{}
\makeatletter
\relax
\def\mn@urlcharsother{\let\do\@makeother \do\$\do\&\do\#\do\^\do\_\do\%\do\~}
\def\mn@doi{\begingroup\mn@urlcharsother \@ifnextchar [ {\mn@doi@} {\mn@doi@[]}}
\def\mn@doi@[#1]#2{\def\@tempa{#1}\ifx\@tempa\@empty \href {http://dx.doi.org/#2} {doi:#2}\else \href {http://dx.doi.org/#2} {#1}\fi \endgroup}
\def\mn@eprint#1#2{\mn@eprint@#1:#2::\@nil}
\def\mn@eprint@arXiv#1{\href {http://arxiv.org/abs/#1} {{\tt arXiv:#1}}}
\def\mn@eprint@dblp#1{\href {http://dblp.uni-trier.de/rec/bibtex/#1.xml} {dblp:#1}}
\def\mn@eprint@#1:#2:#3:#4\@nil{\def\@tempa {#1}\def\@tempb {#2}\def\@tempc {#3}\ifx \@tempc \@empty \let \@tempc \@tempb \let \@tempb \@tempa \fi \ifx \@tempb \@empty \def\@tempb {arXiv}\fi \@ifundefined {mn@eprint@\@tempb}{\@tempb:\@tempc}{\expandafter \expandafter \csname mn@eprint@\@tempb\endcsname \expandafter{\@tempc}}}

\bibitem[\protect\citeauthoryear{{Akeson} et~al.,}{{Akeson} et~al.}{2019}]{2019arXiv190205569A}
{Akeson} R.,  et~al., 2019, \mn@doi [arXiv e-prints] {10.48550/arXiv.1902.05569}, \href {https://ui.adsabs.harvard.edu/abs/2019arXiv190205569A} {p. arXiv:1902.05569}

\bibitem[\protect\citeauthoryear{{Bernstein}}{{Bernstein}}{2010}]{Gary2010}
{Bernstein} G.~M.,  2010, \mn@doi [\mnras] {10.1111/j.1365-2966.2010.16883.x}, \href {https://ui.adsabs.harvard.edu/abs/2010MNRAS.406.2793B} {406, 2793}

\bibitem[\protect\citeauthoryear{{Bernstein} \& {Jarvis}}{{Bernstein} \& {Jarvis}}{2002}]{BernsteinSSSS}
{Bernstein} G.~M.,  {Jarvis} M.,  2002, \mn@doi [\aj] {10.1086/338085}, \href {https://ui.adsabs.harvard.edu/abs/2002AJ....123..583B} {123, 583}

\bibitem[\protect\citeauthoryear{{Bernstein}, {Armstrong}, {Krawiec}  \& {March}}{{Bernstein} et~al.}{2016}]{BFD_Berinstein2016}
{Bernstein} G.~M.,  {Armstrong} R.,  {Krawiec} C.,   {March} M.~C.,  2016, \mn@doi [\mnras] {10.1093/mnras/stw879}, \href {https://ui.adsabs.harvard.edu/abs/2016MNRAS.459.4467B} {459, 4467}

\bibitem[\protect\citeauthoryear{{Bosch} et~al.,}{{Bosch} et~al.}{2018}]{HSCBosch2018}
{Bosch} J.,  et~al., 2018, \mn@doi [\pasj] {10.1093/pasj/psx080}, \href {https://ui.adsabs.harvard.edu/abs/2018PASJ...70S...5B} {70, S5}

\bibitem[\protect\citeauthoryear{{Chang} et~al.,}{{Chang} et~al.}{2013}]{WLsurvey_neffective_Chang2013}
{Chang} C.,  et~al., 2013, \mn@doi [\mnras] {10.1093/mnras/stt1156}, \href {https://ui.adsabs.harvard.edu/abs/2013MNRAS.434.2121C} {434, 2121}

\bibitem[\protect\citeauthoryear{Harris et~al.,}{Harris et~al.}{2020}]{harris2020array}
Harris C.~R.,  et~al., 2020, \mn@doi [Nature] {10.1038/s41586-020-2649-2}, 585, 357

\bibitem[\protect\citeauthoryear{{Heymans} et~al.,}{{Heymans} et~al.}{2006}]{2006MNRAS.368.1323H}
{Heymans} C.,  et~al., 2006, \mn@doi [\mnras] {10.1111/j.1365-2966.2006.10198.x}, \href {https://ui.adsabs.harvard.edu/abs/2006MNRAS.368.1323H} {368, 1323}

\bibitem[\protect\citeauthoryear{{Hirata} \& {Seljak}}{{Hirata} \& {Seljak}}{2003}]{regauss}
{Hirata} C.,  {Seljak} U.,  2003, \mn@doi [\mnras] {10.1046/j.1365-8711.2003.06683.x}, \href {https://ui.adsabs.harvard.edu/abs/2003MNRAS.343..459H} {343, 459}

\bibitem[\protect\citeauthoryear{Hunter}{Hunter}{2007}]{Hunter:2007}
Hunter J.~D.,  2007, \mn@doi [Computing in Science \& Engineering] {10.1109/MCSE.2007.55}, 9, 90

\bibitem[\protect\citeauthoryear{{Huterer}, {Takada}, {Bernstein}  \& {Jain}}{{Huterer} et~al.}{2006}]{2006MNRAS.366..101H}
{Huterer} D.,  {Takada} M.,  {Bernstein} G.,   {Jain} B.,  2006, \mn@doi [\mnras] {10.1111/j.1365-2966.2005.09782.x}, \href {https://ui.adsabs.harvard.edu/abs/2006MNRAS.366..101H} {366, 101}

\bibitem[\protect\citeauthoryear{{Ivezi{\'c}} et~al.,}{{Ivezi{\'c}} et~al.}{2019}]{2019ApJ...873..111I}
{Ivezi{\'c}} {\v{Z}}.,  et~al., 2019, \mn@doi [\apj] {10.3847/1538-4357/ab042c}, \href {https://ui.adsabs.harvard.edu/abs/2019ApJ...873..111I} {873, 111}

\bibitem[\protect\citeauthoryear{{Kaiser}, {Squires}  \& {Broadhurst}}{{Kaiser} et~al.}{1995}]{selection_bias}
{Kaiser} N.,  {Squires} G.,   {Broadhurst} T.,  1995, \mn@doi [\apj] {10.1086/176071}, \href {https://ui.adsabs.harvard.edu/abs/1995ApJ...449..460K} {449, 460}

\bibitem[\protect\citeauthoryear{{Kilbinger}}{{Kilbinger}}{2015}]{kilbinger}
{Kilbinger} M.,  2015, \mn@doi [Reports on Progress in Physics] {10.1088/0034-4885/78/8/086901}, \href {https://ui.adsabs.harvard.edu/abs/2015RPPh...78h6901K} {78, 086901}

\bibitem[\protect\citeauthoryear{{LSST Science Collaboration} et~al.,}{{LSST Science Collaboration} et~al.}{2009}]{2009arXiv0912.0201L}
{LSST Science Collaboration} et~al., 2009, \mn@doi [arXiv e-prints] {10.48550/arXiv.0912.0201}, \href {https://ui.adsabs.harvard.edu/abs/2009arXiv0912.0201L} {p. arXiv:0912.0201}

\bibitem[\protect\citeauthoryear{{Laureijs} et~al.,}{{Laureijs} et~al.}{2011}]{2011arXiv1110.3193L}
{Laureijs} R.,  et~al., 2011, \mn@doi [arXiv e-prints] {10.48550/arXiv.1110.3193}, \href {https://ui.adsabs.harvard.edu/abs/2011arXiv1110.3193L} {p. arXiv:1110.3193}

\bibitem[\protect\citeauthoryear{{Li} \& {Mandelbaum}}{{Li} \& {Mandelbaum}}{2023}]{fpfs}
{Li} X.,  {Mandelbaum} R.,  2023, \mn@doi [\mnras] {10.1093/mnras/stad890}, \href {https://ui.adsabs.harvard.edu/abs/2023MNRAS.521.4904L} {521, 4904}

\bibitem[\protect\citeauthoryear{{Li}, {Katayama}, {Oguri}  \& {More}}{{Li} et~al.}{2018}]{fpfs1}
{Li} X.,  {Katayama} N.,  {Oguri} M.,   {More} S.,  2018, \mn@doi [\mnras] {10.1093/mnras/sty2548}, \href {https://ui.adsabs.harvard.edu/abs/2018MNRAS.481.4445L} {481, 4445}

\bibitem[\protect\citeauthoryear{{Li} et~al.,}{{Li} et~al.}{2022a}]{HSCY3ShapeCatalog}
{Li} X.,  et~al., 2022a, \mn@doi [\pasj] {10.1093/pasj/psac006}, \href {https://ui.adsabs.harvard.edu/abs/2022PASJ...74..421L} {74, 421}

\bibitem[\protect\citeauthoryear{{Li}, {Li}  \& {Massey}}{{Li} et~al.}{2022b}]{fpfs2}
{Li} X.,  {Li} Y.,   {Massey} R.,  2022b, \mn@doi [\mnras] {10.1093/mnras/stac342}, \href {https://ui.adsabs.harvard.edu/abs/2022MNRAS.511.4850L} {511, 4850}

\bibitem[\protect\citeauthoryear{{Li}, {Mandelbaum}  \& {The LSST Dark Energy Science Collaboration}}{{Li} et~al.}{2024a}]{fpfs4}
{Li} X.,  {Mandelbaum} R.,   {The LSST Dark Energy Science Collaboration} 2024a, \mn@doi [arXiv e-prints] {10.48550/arXiv.2408.06337}, \href {https://ui.adsabs.harvard.edu/abs/2024arXiv240806337L} {p. arXiv:2408.06337}

\bibitem[\protect\citeauthoryear{{Li}, {Mandelbaum}, {Jarvis}, {Li}, {Park}  \& {Zhang}}{{Li} et~al.}{2024b}]{fpfs3}
{Li} X.,  {Mandelbaum} R.,  {Jarvis} M.,  {Li} Y.,  {Park} A.,   {Zhang} T.,  2024b, \mn@doi [\mnras] {10.1093/mnras/stad3895}, \href {https://ui.adsabs.harvard.edu/abs/2024MNRAS.52710388L} {527, 10388}

\bibitem[\protect\citeauthoryear{{Liaudat}, {Starck}  \& {Kilbinger}}{{Liaudat} et~al.}{2023}]{Liaudat2023}
{Liaudat} T.~I.,  {Starck} J.-L.,   {Kilbinger} M.,  2023, \mn@doi [Frontiers in Astronomy and Space Sciences] {10.3389/fspas.2023.1158213}, \href {https://ui.adsabs.harvard.edu/abs/2023FrASS..1058213L} {10, 1158213}

\bibitem[\protect\citeauthoryear{{MacCrann} et~al.,}{{MacCrann} et~al.}{2022}]{MacCrann_redshiftdependentshear}
{MacCrann} N.,  et~al., 2022, \mn@doi [\mnras] {10.1093/mnras/stab2870}, \href {https://ui.adsabs.harvard.edu/abs/2022MNRAS.509.3371M} {509, 3371}

\bibitem[\protect\citeauthoryear{{Mandelbaum}}{{Mandelbaum}}{2018}]{2018ARA&A..56..393M}
{Mandelbaum} R.,  2018, \mn@doi [\araa] {10.1146/annurev-astro-081817-051928}, \href {https://ui.adsabs.harvard.edu/abs/2018ARA&A..56..393M} {56, 393}

\bibitem[\protect\citeauthoryear{Mandelbaum, Lackner, Leauthaud  \& Rowe}{Mandelbaum et~al.}{2019}]{mandelbaum_2019_3242143}
Mandelbaum R.,  Lackner C.,  Leauthaud A.,   Rowe B.,  2019, COSMOS real galaxy dataset, \mn@doi{10.5281/zenodo.3242143}, \url {https://doi.org/10.5281/zenodo.3242143}

\bibitem[\protect\citeauthoryear{{Mandelbaum}, {Jarvis}, {Lupton}, {Bosch}, {Kannawadi}, {Murphy}, {Zhang}  \& {LSST Dark Energy Science Collaboration}}{{Mandelbaum} et~al.}{2023}]{2023OJAp....6E...5M}
{Mandelbaum} R.,  {Jarvis} M.,  {Lupton} R.~H.,  {Bosch} J.,  {Kannawadi} A.,  {Murphy} M.~D.,  {Zhang} T.,   {LSST Dark Energy Science Collaboration} 2023, \mn@doi [The Open Journal of Astrophysics] {10.21105/astro.2209.09253}, \href {https://ui.adsabs.harvard.edu/abs/2023OJAp....6E...5M} {6, 5}

\bibitem[\protect\citeauthoryear{{Massey} \& {Refregier}}{{Massey} \& {Refregier}}{2005}]{polar_shapelets_Massey2005}
{Massey} R.,  {Refregier} A.,  2005, \mn@doi [\mnras] {10.1111/j.1365-2966.2005.09453.x}, \href {http://adsabs.harvard.edu/abs/2005MNRAS.363..197M} {363, 197}

\bibitem[\protect\citeauthoryear{{Massey} et~al.,}{{Massey} et~al.}{2007}]{2007MNRAS.376...13M}
{Massey} R.,  et~al., 2007, \mn@doi [\mnras] {10.1111/j.1365-2966.2006.11315.x}, \href {https://ui.adsabs.harvard.edu/abs/2007MNRAS.376...13M} {376, 13}

\bibitem[\protect\citeauthoryear{{Moffat}}{{Moffat}}{1969}]{moffat}
{Moffat} A.~F.~J.,  1969, \aap, \href {https://ui.adsabs.harvard.edu/abs/1969A&A.....3..455M} {3, 455}

\bibitem[\protect\citeauthoryear{{Pujol}, {Kilbinger}, {Sureau}  \& {Bobin}}{{Pujol} et~al.}{2019}]{preciseSim_Pujol2019}
{Pujol} A.,  {Kilbinger} M.,  {Sureau} F.,   {Bobin} J.,  2019, \mn@doi [\aap] {10.1051/0004-6361/201833740}, \href {https://ui.adsabs.harvard.edu/abs/2019A\&A...621A...2P} {621, A2}

\bibitem[\protect\citeauthoryear{{Refregier}}{{Refregier}}{2003}]{shapeletsI_Refregier2003}
{Refregier} A.,  2003, \mn@doi [\mnras] {10.1046/j.1365-8711.2003.05901.x}, \href {http://adsabs.harvard.edu/abs/2003MNRAS.338...35R} {338, 35}

\bibitem[\protect\citeauthoryear{{Refregier}, {Kacprzak}, {Amara}, {Bridle}  \& {Rowe}}{{Refregier} et~al.}{2012}]{image_noise_bias}
{Refregier} A.,  {Kacprzak} T.,  {Amara} A.,  {Bridle} S.,   {Rowe} B.,  2012, \mn@doi [\mnras] {10.1111/j.1365-2966.2012.21483.x}, \href {https://ui.adsabs.harvard.edu/abs/2012MNRAS.425.1951R} {425, 1951}

\bibitem[\protect\citeauthoryear{{Rowe}, {Jarvis}  \& {Mandelbaum}}{{Rowe} et~al.}{2014}]{galsim}
{Rowe} B.,  {Jarvis} M.,   {Mandelbaum} R.,  2014, {GalSim: Modular galaxy image simulation toolkit}, Astrophysics Source Code Library, record ascl:1402.009 (\mn@eprint {ascl} {1402.009})

\bibitem[\protect\citeauthoryear{{Sheldon} \& {Huff}}{{Sheldon} \& {Huff}}{2017}]{Metacal2017}
{Sheldon} E.~S.,  {Huff} E.~M.,  2017, \mn@doi [\apj] {10.3847/1538-4357/aa704b}, \href {https://ui.adsabs.harvard.edu/abs/2017ApJ...841...24S} {841, 24}

\bibitem[\protect\citeauthoryear{{Sheldon}, {Becker}, {MacCrann}  \& {Jarvis}}{{Sheldon} et~al.}{2020}]{metaDet_Sheldon2020}
{Sheldon} E.~S.,  {Becker} M.~R.,  {MacCrann} N.,   {Jarvis} M.,  2020, \mn@doi [\apj] {10.3847/1538-4357/abb595}, \href {https://ui.adsabs.harvard.edu/abs/2020ApJ...902..138S} {902, 138}

\bibitem[\protect\citeauthoryear{{Sheldon}, {Becker}, {Jarvis}, {Armstrong}  \& {LSST Dark Energy Science Collaboration}}{{Sheldon} et~al.}{2023}]{metaDet_LSST2023}
{Sheldon} E.~S.,  {Becker} M.~R.,  {Jarvis} M.,  {Armstrong} R.,   {LSST Dark Energy Science Collaboration} 2023, \mn@doi [The Open Journal of Astrophysics] {10.21105/astro.2303.03947}, \href {https://ui.adsabs.harvard.edu/abs/2023OJAp....6E..17S} {6, 17}

\bibitem[\protect\citeauthoryear{{The LSST Dark Energy Science Collaboration} et~al.,}{{The LSST Dark Energy Science Collaboration} et~al.}{2018}]{2018arXiv180901669T}
{The LSST Dark Energy Science Collaboration} et~al., 2018, \mn@doi [arXiv e-prints] {10.48550/arXiv.1809.01669}, \href {https://ui.adsabs.harvard.edu/abs/2018arXiv180901669T} {p. arXiv:1809.01669}

\bibitem[\protect\citeauthoryear{Virtanen et~al.,}{Virtanen et~al.}{2020}]{2020SciPy-NMeth}
Virtanen P.,  et~al., 2020, \mn@doi [Nature Methods] {10.1038/s41592-019-0686-2}, \href {https://rdcu.be/b08Wh} {17, 261}

\bibitem[\protect\citeauthoryear{{Zhang} et~al.,}{{Zhang} et~al.}{2023}]{Tianqing_GeneralFrameworkPSF}
{Zhang} T.,  et~al., 2023, \mn@doi [\mnras] {10.1093/mnras/stad1801}, \href {https://ui.adsabs.harvard.edu/abs/2023MNRAS.525.2441Z} {525, 2441}

\makeatother
\end{thebibliography}




\appendix

\section{Shapelet Moments of Gaussian Profiles}
\label{sec:apA}
In this appendix, we provide intuition into the additional information captured by non-adaptive fourth order moment measurements of a Gaussian profile under shear and compare it against the second order moment measurement.

Let $f(\bm{x}, \sigma_g)$ be a Gaussian profile with scale radius $\sigma_g$, which corresponds to the shapelet basis $\chi_{00}(\bm{x}| \sigma_g)$. Under shear, the profile transforms as:
\begin{align}
    \chi_{00}(\bm{x}|\sigma_g) \to \chi_{00}(\bm{x}|\sigma_g) &= \bar{\chi}_{00}(\bm{x}|\sigma_g)\\
    &\nonumber- \frac{\sqrt{2}\gamma_1}{2}\left(\chi_{2,-2}(\bm{x}|\sigma_g) + \chi_{2,2}(\bm{x}|\sigma_g)\right)\\
    &\nonumber+ \frac{\sqrt{2}\gamma_2}{2}\left(\chi_{2,-2}(\bm{x}|\sigma_g) - \chi_{2,2}(\bm{x}|\sigma_g)\right),
\end{align}
where $\bar{\chi}_{00}$ is the intrinsic shapelet basis in the absence of shear. The corresponding shapelet moments $M_{nm}$ of the Gaussian profile can be estimated by projecting the profile to shapelet bases using equation~\eqref{eq:shapelets_modes}.

\begin{figure}
    \centering
    \includegraphics[width=0.5\textwidth]{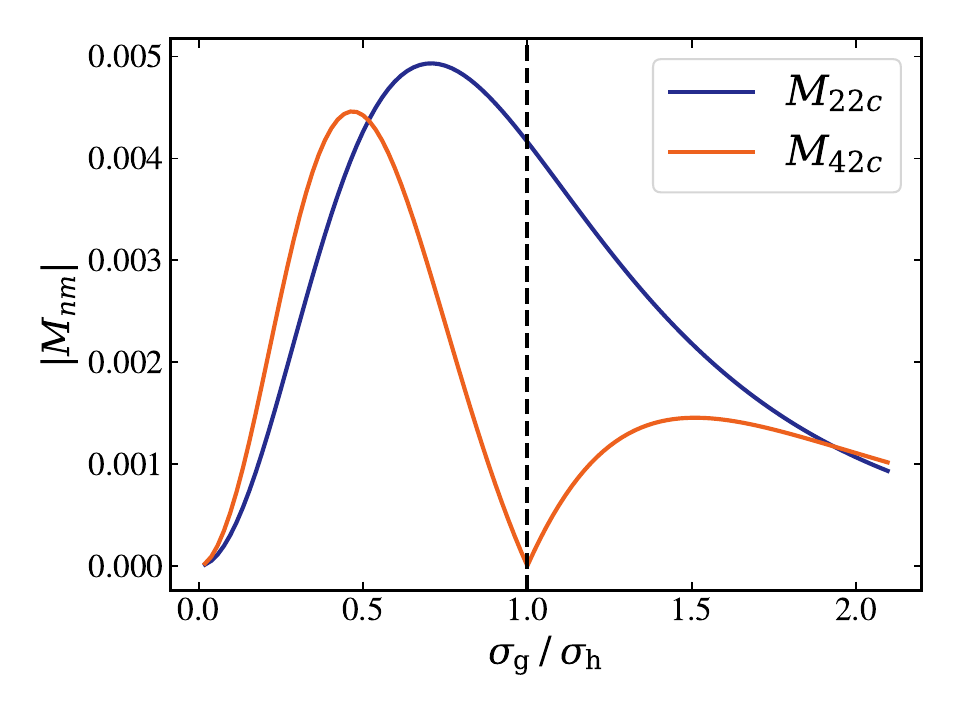}
    \caption{Spin-2 component of the second (blue) and fourth-order (orange) shapelet moments of a Gaussian profile with input shear $\gamma_1=0.02$ as a function of the ratio between the scale of the Gaussian profile to the shapelet kernel. The vertical dashed line indicates when the two scales are identical. 
    }
    \label{fig:gaussianprofile}
\end{figure}

The generalized Laguerre polynomials are orthogonal over $[0, \infty]$ with a weighting function $x^\alpha e^{-x}$:
\begin{equation}
    \label{eq:ortho}
    \int_0^\infty x^\alpha e^{-x} L_n^{(\alpha)}(x) L_m^{(\alpha)}(x) dx = \frac{\Gamma(n+\alpha+1)}{n!} \delta_{n,m},
\end{equation}
where $\delta_{n,m}$ is the Kronecker delta function. Using the orthogonality relation in equation~\eqref{eq:ortho}, if the scale of shapelets and Gaussian is the same ($\sigma_h = \sigma_g$), only the shapelet moments $M_{00}, M_{2,-2}$, and $M_{22}$ are nonzero and fourth-order moments are zero, i.e. $M_{40} = M_{42} = 0$. However, in most cases, $\sigma_h \neq \sigma_g$ or $M_{42} \neq 0$, so shear can be estimated using fourth-order moments. In Fig.~\ref{fig:gaussianprofile}, we show the spin-2 component of the second-and fourth-order shapelet moments and show that the fourth-order moment is 0 when the scale of the shapelet kernel and Gaussian is the same. For shapelet kernel scale that is smaller than half of the scale of the Gaussian profile (i.e., ratio smaller than 0.5), by comparing the magnitude of each shapelet moment, we find that the fourth order moments have more information than the second moment as they can probe smaller radii than the second moment. It is also worth noting that by comparing the magnitude of the shapelet moment, the fourth order moment has more information on small scales compared to large scales.  


\bsp	
\label{lastpage}
\end{document}